\documentclass[conference]{IEEEtran}

\ifCLASSINFOpdf
\else
\fi

\let\labelindent\relax

\usepackage{epsfig,amsmath,amsfonts,multirow,graphicx,makecell,caption,soul,csquotes,color,wrapfig,subcaption,mathtools,bm,spverbatim,booktabs,xcolor,color,amsthm,tcolorbox,wrapfig,enumitem}
\usepackage[e]{esvect}
\usepackage{graphics}
\usepackage{marvosym,listings,etoolbox}
\usepackage{adjustbox}
\usepackage[space]{cite}
\usepackage{subcaption}
\DeclareCaptionSubType * [alph]{table}
\usepackage{multicol}
\usepackage{lipsum}

\usepackage{bbding}

\usepackage{array}
\usepackage{longtable}
\usepackage{colortab}
\usepackage{colortbl}
\usepackage{arydshln}

\makeatletter
\renewcommand\p@subfigure{\thefigure\,}

\makeatother

\usepackage{fancyhdr}

\fancypagestyle{empty}{\fancyfoot[C]{\vspace*{-1.8\baselineskip}\thepage}}
\fancypagestyle{plain}{\fancyfoot[C]{\vspace*{-1.8\baselineskip}\thepage}}

\usepackage[first=0,last=9]{lcg}
\usepackage{colortbl}
\definecolor{Gray}{gray}{0.8}

\usepackage{url}

\usepackage{hyperref}
\usepackage{xurl}

\def\tablename{Table}
\def\figurename{Figure}

\makeatletter
\newif\if@restonecol
\makeatother

\usepackage[boxed, ruled, vlined, linesnumbered]{algorithm2e}
\SetKwRepeat{Do}{do}{while}

\newenvironment{changemargin}[2]{\begin{list}{}{
	\setlength{\topsep}{0pt}\setlength{\leftmargin}{0pt}
	\setlength{\rightmargin}{0pt}
	\setlength{\listparindent}{\parindent}
	\setlength{\itemindent}{\parindent}
	\setlength{\parsep}{0pt plus 1pt}
	\addtolength{\leftmargin}{#1}\addtolength{\rightmargin}{#2}
	}\item}
	{\end{list}}

\usepackage{xr}

\makeatletter
\newcommand*{\addFileDependency}[1]{%
  \typeout{(#1)}
  \@addtofilelist{#1}
  \IfFileExists{#1}{}{\typeout{No file #1.}}
}
\makeatother

\usepackage{scalerel}[2016/12/29]

\usepackage{diagbox}

\usepackage{array}
\newcolumntype{P}[1]{>{\centering\arraybackslash}p{#1}}

\DeclareMathAlphabet\mathbfcal{OMS}{cmsy}{b}{n}

\iffalse

\newcommand{\nsection}[1]{\section{#1}}
\newcommand{\nsubsection}[1]{\subsection{#1}}
\newcommand{\nsubsubsection}[1]{\subsubsection{#1}}

\else

\newcommand{\nsection}[1]{\vspace{-0.05in}\section{#1} \vspace{-0.03in}}
\newcommand{\nsubsection}[1]{\vspace{-0.04in}\subsection{#1}\vspace{-0.03in}}
\newcommand{\nsubsubsection}[1]{\vspace{-0.01in}\subsubsection{#1}\vspace{-0.03in}}

\fi

\usepackage{mdframed}
\mdfdefinestyle{rebuttalstyle}{innerleftmargin=3pt, innerrightmargin=3pt, innertopmargin=3pt, innerbottommargin=3pt}

\newcounter{response}[section]

\newcounter{revision}[section]

\newcounter{link}[section]

\newcounter{comments}[section]

\newcounter{observation} %
\newenvironment{observation}[1]{\refstepcounter{observation}
\vspace{-1mm}
\begin{mdframed}[style=rebuttalstyle]
\noindent \textbf{Finding~\theobservation} (#1): \itshape
}
{
\end{mdframed}
\vspace{-0.15mm}
}

\definecolor{gray}{rgb}{0.7,0.7,0.7}

\newcommand{\cut}[1]{}

\newcommand{\newpart}[1]{#1}

\newcommand{\DemoWeb}{\textcolor{blue}{\textbf{\url{https://sites.google.com/view/cav-sec/new-gen-lidar-sec}}}}

\newcommand{\deleted}[1]{}

\newcommand{\circled}[1]{\raisebox{.5pt}{\textcircled{\raisebox{-.9pt} {{\small #1}}}}}
\makeatletter
\newcommand\notsotiny{\@setfontsize\notsotiny{7}{8}}
\makeatother

\usepackage{lipsum}

\begin{document}
\title{
LiDAR Spoofing Meets the New-Gen: \\ Capability Improvements, Broken Assumptions, and New Attack Strategies}

\author{\IEEEauthorblockN{
Takami Sato\IEEEauthorrefmark{1}\IEEEauthorrefmark{2}\thanks{\IEEEauthorrefmark{1}co-first authors},
Yuki Hayakawa\IEEEauthorrefmark{1}\IEEEauthorrefmark{3},
Ryo Suzuki\IEEEauthorrefmark{1}\IEEEauthorrefmark{3}, 
Yohsuke Shiiki\IEEEauthorrefmark{1}\IEEEauthorrefmark{3},
Kentaro Yoshioka\IEEEauthorrefmark{3}, 
Qi Alfred Chen\IEEEauthorrefmark{2}}
\IEEEauthorblockA{\IEEEauthorrefmark{2}University of California, Irvine, Department of Computer Science\\
\IEEEauthorrefmark{3}Keio University, Department of Electronics and Electrical Engineering
}
\IEEEauthorblockA{
\IEEEauthorrefmark{2}\{takamis, alfchen\}@uci.edu, \IEEEauthorrefmark{3}\{hykwyuk, suzuki.ryo, 
kyoshioka47\}@keio.jp,  \IEEEauthorrefmark{3}shiiki@iskr.elec.keio.ac.jp
}
}

\IEEEoverridecommandlockouts
\makeatletter\def\@IEEEpubidpullup{5\baselineskip}\makeatother
\IEEEpubid{\parbox{\columnwidth}{
    Network and Distributed System Security (NDSS) Symposium 2024\\
    26 February - 1 March 2024, San Diego, CA, USA\\
    ISBN 1-891562-93-2\\
    https://dx.doi.org/10.14722/ndss.2024.23350\\
    www.ndss-symposium.org
}
\hspace{\columnsep}\makebox[\columnwidth]{}}

\maketitle

\begin{abstract}
LiDAR (Light Detection And Ranging) is an indispensable sensor for precise long- and wide-range 3D sensing, which directly benefited the recent rapid deployment of autonomous driving (AD). Meanwhile, such a safety-critical application strongly motivates its security research. 
A recent line of research finds that one can manipulate the LiDAR point cloud and fool object detectors by firing malicious lasers against LiDAR. However, these efforts face 3 critical research gaps: (1) considering only one specific LiDAR (VLP-16); (2) assuming unvalidated attack capabilities; and (3) evaluating object detectors with limited spoofing capability modeling and setup diversity.

To fill these critical research gaps, we conduct the first large-scale measurement study on LiDAR spoofing attack capabilities on object detectors with 9 popular LiDARs, covering both first- and new-generation LiDARs, and 3 major types of object detectors trained on 5 different datasets. To facilitate the measurements, we (1) identify spoofer improvements that significantly improve the latest spoofing capability, (2) identify a new object removal attack that overcomes the applicability limitation of the latest method to new-generation LiDARs, and (3) perform novel mathematical modeling for both object injection and removal attacks based on our measurement results. Through this study, we are able to uncover a total of 15 novel findings, including not only completely new ones due to the measurement angle novelty, but also many that can directly challenge the latest understandings in this problem space. We also discuss defenses.

\end{abstract}

\nsection{Introduction} \label{sec:intro}

LiDAR (Light Detection And Ranging)
is one of the most innovative sensors in the past decade. By shooting a laser pulse and measuring its reflection,
LiDAR can provide a detailed 3D understanding of the surrounding environment.
Autonomous Driving (AD) is one of the most benefited applications of the high-speed and high-precision sensing of LiDARs. After LiDAR showed its effectiveness in the 2007 DARPA Urban Challenge\cite{urmson2007tartan}, it has been widely recognized as an essential sensor for Level-4 AD and has been adopted in almost all recent robotaxi services (Waymo One~\cite{waymoone}, Cruise~\cite{Cruise}) and AD vehicles operating in the US~\cite{Motional, Nuro}.
While highly beneficial to our everyday life and society, AD is also highly security-critical as even a small operational error can cause fatal consequences~\cite{ubersafety}. To address this, numerous researchers have been conducting security analyses on LiDARs~\cite{petit2015remote, shin2017illusion, cao2019adversarial, jiachen2020towards, cao2021invisible, hallyburton2022security, cao2023you} due to their critical role in AD perception. The major security concern of LiDARs is the fundamental vulnerability against malicious laser shooting, or \textit{LiDAR spoofing attacks}.
The recent research along this line~\cite{jiachen2020towards, zhongyuan2021object, hallyburton2022security} found that such attacks can cause both false positives (injecting a non-existing fake object) and false negatives (removing an existing object). However, we find that there are 3 critical research gaps in these prior efforts:

\begin{figure}[t!]
\centering
\includegraphics[width=\linewidth]{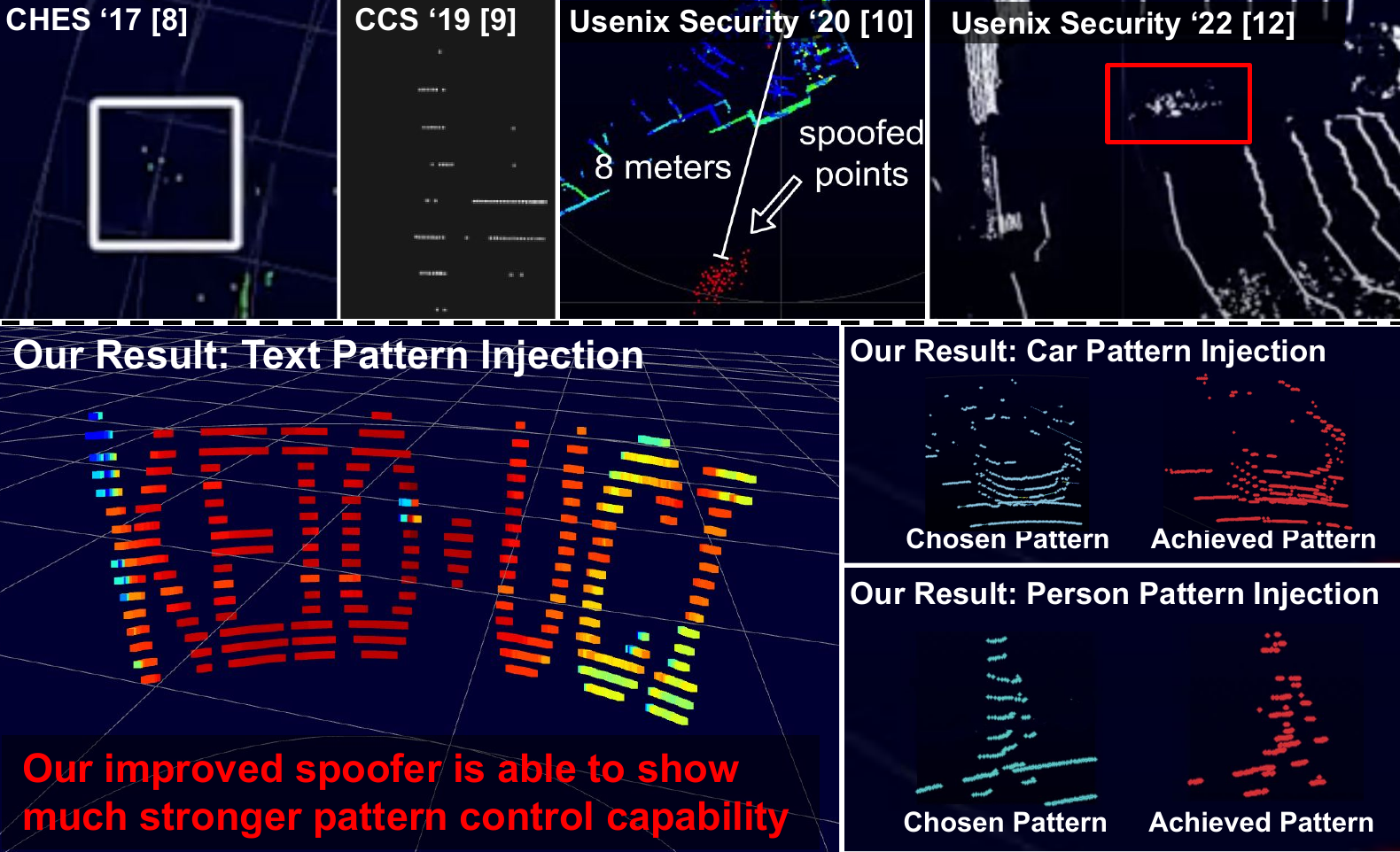}
\caption{Demonstration of the Chosen Pattern Injection (CPI) attack capability with $>$6,000 spoofed points with our improved LiDAR spoofer. This significantly improves the spoofing attack capability from prior works:~\cite{shin2017illusion} ($\sim$10 points),~\cite{cao2019adversarial} ($\sim$20 points)~\cite{jiachen2020towards} ($\sim$200 points), and~\cite{hallyburton2022security} ($\sim$200 points).
}
\label{fig:arbitrary_point}
\end{figure}

\textbf{Considering only one specific LiDAR:}
Velodyne VLP-16~\cite{VLP16} has been dominantly used in the prior works since it is viewed as a \textit{de facto} choice for LiDAR spoofing evaluation after the first practical spoofing attack was proposed in 2017~\cite{shin2017illusion}. The following works thus all evaluate their attacks only on VLP-16~\cite{cao2019adversarial, jiachen2020towards, hallyburton2022security, cao2023you} or use the attack capability on VLP-16 to justify the validity of their threat model~\cite{zhongyuan2021object, hau2021shadow}. 
Although these results are valid on VLP-16, there is no guarantee that these results are still valid in more recent LiDARs, known as next-generation or new-generation (or \textit{new-gen}) LiDARs~\cite{yoshioka2022tutorial}, as opposed to the first-generation (or \textit{first-gen}) ones such as VLP-16. The new-gen LiDARs have more advanced security-related features, such as laser timing randomization and pulse fingerprinting. 
Prior works~\cite{cao2019adversarial, cao2023you, shin2017illusion} partially discussed some of them as potential defenses, but none of them experimentally studied their impact and effectiveness against LiDAR spoofing.

\textbf{Assuming unvalidated attack capabilities:}
So far, all prior works on fake object injection~\cite{cao2019adversarial, jiachen2020towards, hallyburton2022security} assume a \textit{Chosen Pattern Injection (CPI)} attack capability, i.e., an attacker can spoof a specifically chosen point cloud pattern that was carefully optimized/identified offline beforehand.
For example, the Adv-LiDAR attack~\cite{cao2019adversarial} is a white-box adversarial attack that needs a specific pattern to trigger a vulnerability in deep neural networks (DNNs). The black-box attacks in~\cite{jiachen2020towards, hallyburton2022security} also require a specific pattern in the shape of a vehicle surface.
However, none of them have systematically
studied how achievable this is in practice.
The top area of Fig.~\ref{fig:arbitrary_point} shows the demonstrated spoofing capabilities so far from prior works~\cite{shin2017illusion, cao2019adversarial, jiachen2020towards, hallyburton2022security}. As shown, none of them were able to show strong pattern control capabilities, not to mention to provide a systematic quantification of such a capability to ensure valid security analysis on the object detector side.

\textbf{Evaluating object detectors with limited spoofing capability modeling and setup diversity:} To study the spoofing attack impacts on the object detector side, prior works typically leverage a mathematical modeling of the spoofing capability to achieve high analysis scalability and flexibility~\cite{cao2019adversarial,jiachen2020towards, hallyburton2022security}. However, the modelings used so far not only fail to correctly capture real-world attack characteristics such as spoofing inaccuracies (detailed in~\S\ref{sec:cpi_capability}, which can significantly bias the analysis results as shown in~\S\ref{sec:od_injection_eval}), but also lack the coverage of more recent LiDAR features that can significantly change the spoofing capabilities (\S\ref{sec:sec_enchance_feats}). Moreover, the object detector model setup is also limited, e.g., no prior works have evaluated the same model design trained on different training datasets, which we found can significantly change the model robustness analysis results (\S\ref{sec:od_injection_eval} and~\S\ref{sec:od_removal}).

To fill these critical research gaps, we conduct the first large-scale measurement study on LiDAR spoofing attack capabilities against object detection. Specifically, our study is driven by 3 key novel research questions (RQ): \\
\textbf{RQ1}: \textit{Are the common design-level assumptions made in prior LiDAR spoofing attacks actually realizable? If so, can they also hold for the more recent new-gen LiDARs?} \\
\textbf{RQ2}: \textit{Do different types of LiDARs, especially the new-gen ones with security-related features, have different vulnerability characteristics to LiDAR spoofing attacks?} \\
\textbf{RQ3}: \textit{Does the vulnerability status of popular object detectors to spoofing attacks significantly change due to the new-gen LiDAR features and different model setups?}

To comprehensively address these RQs, we devote significant engineering efforts to reproduce and/or improve the current state-of-the-art attacks: For RQ1, to properly explore the potential to achieve the CPI attack capability, we improve the spoofing device on their optical and electronics parts (\S\ref{sec:optical_setup}), which enables us to be the first to demonstrate and quantify the CPI attack capability, which is commonly assumed but never clearly demonstrated in prior works~\cite{cao2019adversarial, jiachen2020towards, cao2023you, hallyburton2022security, zhongyuan2021object} and shown in Fig.~\ref{fig:arbitrary_point}. 
For RQ2, to adequately measure the vulnerability status of new-gen LiDARs, we explore \textit{asynchronized} (\S\ref{sec:lidar_spoofing_attack}) spoofing attacks since synchronized ones are directly foiled by their common security-related features such as timing randomization and pulse fingerprinting (\S\ref{sec:inj_other_lidars}). However, since all existing works only consider first-gen LiDARs, their designs predominately focus on synchronized attacks~\cite{shin2017illusion, cao2019adversarial, jiachen2020towards, hallyburton2022security, cao2023you}, leaving the asynchronized attack design space under-explored. 
For RQ3, to enable large-scale evaluation of LiDAR spoofing attack capabilities against different object detectors, we perform novel mathematical modeling of the spoofing attack capabilities on different types of LiDARs based on our measurements for RQ1 and RQ2, which is not only the first to perform such modelling for first- and new-gen LiDARs, but also the first to model for object removal attacks.

In the measurement study, we cover major types of LiDAR spoofing attacks against (1) 9 popular LiDAR models in total, covering both the classic first-gen ones (e.g., VLP-16) and the new-gen ones with new security-related features; and (2) 3 major types of object detectors trained on 5 different datasets. To the best of our knowledge, this is the first large-scale measurement study on this topic in terms of \textit{LiDAR numbers} (none of the prior works studied over 1 LiDAR model, while we study 9), \textit{LiDAR types} (all prior works concentrate on first-gen ones, and we are the first to study the new-gen ones), and also \textit{training datasets} (no prior works studied the same model design with over 1 dataset, while we study 5).

Through this study, we are able to identify a total of \textit{15 novel findings}, which include not only completely new ones due to the measurement angle novelty (e.g., measurements on new-gen LiDARs), but also many that directly challenge the latest understandings in this problem space. For example:

\begin{itemize}[leftmargin=0.2in]

\item We find that with more careful spoofer optics and electronics implementations, an attacker actually does not really need to exploit object detector-level vulnerabilities to achieve a near-front road object injection, which is a common assumption made in prior works~\cite{cao2019adversarial, jiachen2020towards, hallyburton2022security};

\item We find that VLP-16 is actually the \textit{only} LiDAR model for which the CPI attack capability assumption is feasible, which is another key design assumption made in latest prior works~\cite{cao2019adversarial, jiachen2020towards, hallyburton2022security};

\item We find that the new-gen LiDAR features previously expected to be capable of largely mitigating or even preventing spoofing (e.g., timing randomization~\cite{cao2019adversarial, cao2023you}, pulse fingerprinting~\cite{shin2017illusion}) may not be effective as expected 
(\S\ref{sec:sec_enchance_feats}, \S\ref{sec:od_injection_eval});

\item We find that the latest synchronized object removal attack can no longer be applied to the new-gen LiDARs, but there exist asynchronized object removal attacks that can overcome such a limitation and can lead to a similar level of practical attack capabilities (\S\ref{sec:removal_attack}).

\end{itemize}

In summary, our study has the following contributions:
\begin{itemize}[leftmargin=0.2in]
    \item We conduct the first large-scale measurement study on LiDAR spoofing attack capabilities on object detectors with 9 LiDARs, covering both first- and new-generation ones, and 3 major types of object detectors. We are also the first to investigate the impacts of new security-related features in more recent LiDARs from the security perspective. 
    \item To facilitate the measurements, we not only identify spoofer improvements that significantly improve the latest spoofing capability, but also identify  a new asynchronized object removal attack that overcomes the applicability limitation of the latest method to new-gen LiDARs while having a similar level of practical attack capability.
   
    \item To facilitate the object detector-level measurements, we perform novel mathematical modeling of the spoofing capabilities for both object injection and removal attacks based on our measurement results, which is the first to model for both first- and new-gen LiDARs and also to model for object removal attacks.
    \item Our study is able to uncover a total of 15 novel findings, including not only completely new ones due to the measurement angle novelty, but also many that can directly challenge the latest understandings in this problem space.
\end{itemize}

\textbf{Data/code release.} All data, code, and hardware designs (e.g., for the new spoofer) are released at \DemoWeb.

\nsection{Background and Related Works} \label{sec:background}
\vspace{0.1in}

\nsubsection{LiDAR Basics and Recent Trend}
\label{sec:lidar_basic}
\label{sec:lidar_tax}
LiDAR is an active sensor that can obtain high-resolution 3D point cloud data of the surrounding environment. The most common type of LiDAR today is direct time-of-flight (dToF) LiDAR, which fires laser pulses to the environment and uses their reflections to actively measure (or ``scan'') the surrounding objects. Specifically, for each laser pulse, it uses the time difference between the firing and reflection to obtain the 3D position of a ``point'' on the object surface. By performing laser firing to different horizontal angles (\textit{azimuth}) and vertical angles (\textit{altitude}), the point measurements thus form a ``point cloud'' (Fig.~\ref{fig:arbitrary_point}). Such LiDARs typically use laser peak power (e.g., $\sim$10 W) significantly higher than sunlight even after a long flight, which allows them to detect objects more than 200 meters away even under high ambient lights.

\textbf{Recent Trend: New-Gen LiDARs.} Early Velodyne LiDARs such as VLP-16~\cite{VLP16} and VLP-32c~\cite{VLP32c} are commonly known as the first-generation (\textit{first-gen}) LiDARs~\cite{yoshioka2022tutorial}, which naively integrate the classic point-wise laser ranging systems. The advent of first-gen LiDARs has significantly boosted critical real-world applications such as autonomous driving (AD), but its complex mechanical design increases costs and limits scalability.
To overcome the limitations, the new-generation (\textit{new-gen}) LiDARs~\cite{yoshioka2022tutorial} mount all components, such as the photodetector and the readout circuitry, on a single chip (called a system-on-chip (SoC) approach). This not only reduces the cost and improves the scalability of the system, but also allows new designs of laser firing and receiving.  %
For example, microelectromechanical systems (MEMS) LiDAR~\cite{wang2020mems} move a mirror to scan a wide azimuth range instead of mechanical rotation. Flash LiDARs~\cite{roriz2021automotive, ousterflash} fire a broad laser that covers the entire field of view (FOV) and calculates the distance by compensating its weak return laser by accumulation. 
Moreover, the SoC approach allows more complex signal processing, such as a large number of simultaneous laser firings~\cite{OS1-32}, laser timing randomization~\cite{OS1-32, Helios}, and fingerprinting~\cite{XT32}, to be robust against challenging environments, e.g., multiple LiDARs operating adjacent to each other.

All in all, the recent new-gen LiDARs substantially improved the electronics and the scanning mechanisms even though they still have the same basic design: laser firing and receiving. However, none of the prior works on LiDAR spoofing attacks have studied the security property of such new-gen LiDARs; \textit{all} of them are performed on only the first-gen rotational ones, with a predominate focus on in fact \textit{only 1 specific LiDAR model}: VLP-16~\cite{shin2017illusion, cao2019adversarial, jiachen2020towards, hallyburton2022security, cao2023you}. The major design differences between first- and new-gen LiDARs are likely to cause significant differences in their security characteristics, which thus motivates this study.

\nsubsection{Object Detection on Point Cloud} \label{sec:3d_obj}

In real-world applications such as autonomous driving (AD), object detection is no doubt one of the most critical roles for LiDAR. As in other computer vision areas, 3D object detection on point cloud is substantially benefited by the recent progress of DNN. However, traditional DNN architectures such as CNN cannot be directly applied due to the irregular and unordered structure of point clouds~\cite{li2018pointcnn}. To handle such data complexity, 3 major types of 3D object detection methods are widely adopted: \textit{voxel-based}, \textit{point-based}, and \textit{point voxel-based} methods~\cite{qian2022object}. More details are in Appendix~\ref{appndix:obj_detector}.

\nsubsection{LiDAR Spoofing Attacks against Object Detection} \label{sec:lidar_spoofing_attack}

Due to the basic sensing mechanism of laser firing and receiving, LiDARs are fundamentally vulnerable to malicious laser shooting, or \textit{LiDAR spoofing attacks}.
Specifically, such attacks work by using an external attack device (``spoofer'') to fire laser pulses back to the victim LiDAR in order to manipulate the time measurements of the laser-receiving events and thus the corresponding 3D position measurements (\S\ref{sec:lidar_basic}). Such point position manipulations can thus be strategically used to achieve attack effects on the 3D object detector side, more specifically (1) \textit{Object injection}, e.g., by moving the positions of a set of points to form a 3D pattern to cause a false-positive detection of a non-existing object; and (2) \textit{Object removal}, e.g., by moving the points originally on an object to elsewhere to cause false-negative detection of such an object.

\nsubsubsection{Attack Taxonomy}
\label{sec:attack_taxonomy}

\begin{figure}[t!]
\centering
\includegraphics[width=\linewidth]{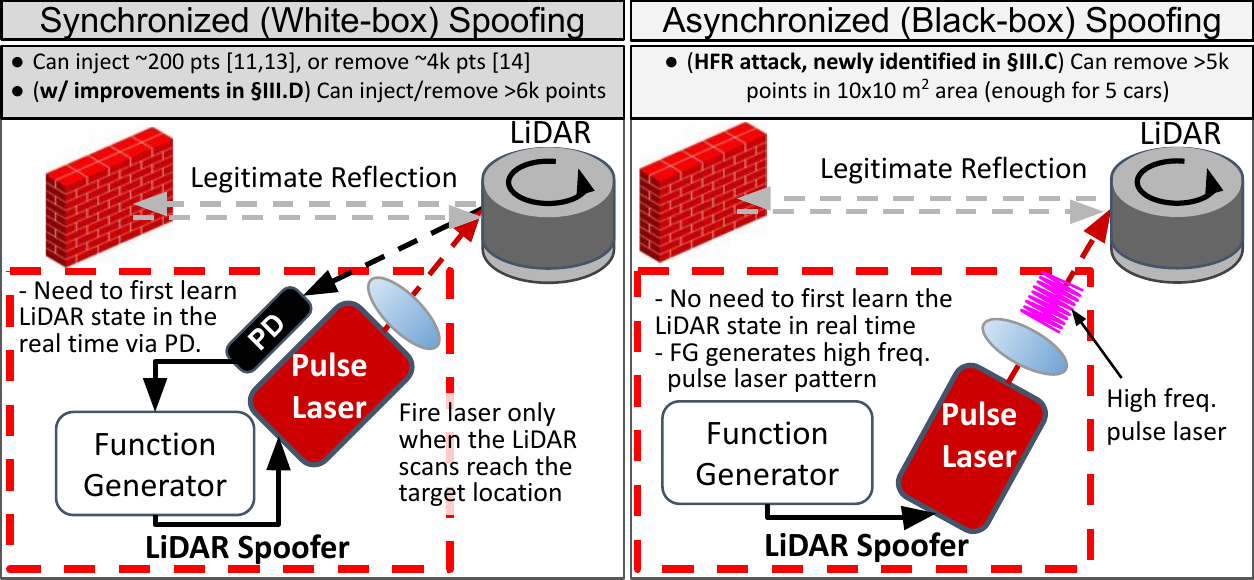}
\caption{
Illustration of the synchronized and asynchronized LiDAR spoofing techniques with the latest attack capabilities. Synchronized spoofing needs white-box knowledge of the victim LiDAR scanning patterns and an extra device (Photodetector, or PD) for synchronization (\S\ref{sec:lidar_spoofing_attack}), while asynchronized spoofing does not need these (i.e., black-box LiDAR attack).
}
\label{fig:sync_and_async_attack}
\end{figure}

\begin{table}[t!]
\centering
\footnotesize
\caption{Taxonomy of existing LiDAR spoofing attacks against 3D object detection. Sync. and Async. refer to synchronized and asynchronized spoofing techniques (\S\ref{sec:lidar_spoofing_attack}).}
\label{tbl:tax_attacks}
\setlength{\tabcolsep}{5.6pt}
\renewcommand{\arraystretch}{0.9}
\begin{tabular}{ccc}
\toprule
Spoofing & \multicolumn{2}{c}{Object Detection Model-Level Attack Effect} \\\cline{2-3} 
Technique & Object Injection & Object Removal \\\hline
\begin{tabular}[c]{@{}l@{}}\ \ \ \ \ Sync.\\(White-box)\end{tabular}   & 
\begin{tabular}[c]{@{}l@{}}\hspace{3em}Adv-LiDAR~\cite{cao2019adversarial},\\ Occlusion~\cite{jiachen2020towards}, Frustum~\cite{hallyburton2022security}\end{tabular}
        & \hspace{0.5em}PRA~\cite{cao2023you}, ORA~\cite{petit2015remote}       \\
\hdashline
\begin{tabular}[c]{@{}l@{}}\ \ \ \ Async.\\(Black-box)\end{tabular}  & Relay$^{*}$~\cite{petit2015remote}      & \begin{tabular}[c]{@{}l@{}}\hspace{2em}Saturating$^{\dag}$~\cite{shin2017illusion},\\ \hspace{2em}{HFR (\S\ref{sec:high_freq_attack})}\end{tabular}           \\
 \toprule
\end{tabular}
\raggedright

$^{*}$ Cannot inject fake objects (1) closer than the spoofer itself, and (2) in a different angle towards the victim than the spoofer itself.\\
$^{\dag}$ Only show removal of a 41$\times$42 cm$^2$ metal plate in short duration (4 sec)
\end{table}
Table~\ref{tbl:tax_attacks} shows a taxonomy of existing LiDAR spoofing attacks on object detection based on the LiDAR spoofing technique (specifically, the requirement of \textit{synchronization}) and the targeted object detector-level attack effect.
\textit{Synchronization} means to synchronize the malicious laser firing timing with the victim LiDAR scanning (i.e., laser firing) timing, which can enable precise control of the attack laser-receiving timing and thus the corresponding positions of the spoofed points. 
Fig.~\ref{fig:sync_and_async_attack} illustrates the synchronized and asynchronized spoofing techniques with the latest attack capabilities. As shown, to achieve synchronization, the attacker needs to use an extra device (photodetector, or PD, in Fig.~\ref{fig:sync_and_async_attack}) to first learn the current state of the victim LiDAR scanning in the real time, and then use the victim LiDAR's scanning pattern to derive future victim laser-firing timings for synchronized attack laser-firing. More detailed explanations are in Appendix~\ref{appndix:sync_attack}. This process requires precise and predictable knowledge of the victim LiDAR's scanning pattern beforehand, which thus can be viewed as a \textit{white-box} LiDAR attack. Those that do not assume synchronization (i.e., \textit{asynchronized} attacks) can be applied without knowing anything about the LiDAR internal scanning logic, which can be viewed as \textit{black-box} attacks.

\textbf{Spoofing Attacks for Object Injection.} 
The very first LiDAR spoofing attack was discovered in 2015~\cite{petit2015remote}, which found that shooting lasers to a victim LiDAR can inject $\geq$200 points and achieve fake object injection at the object detector side. This attack works by relaying the laser received from the victim LiDAR, which thus does not require synchronization. However, this also makes it impossible to inject objects (1) in closer positions than the spoofer itself, and (2) in a different angle towards the victim than the spoofer. These limitations make it not very useful in real-world attack scenarios since at the time of spoofing, the attackers would hope that the most threatening object to the victim vehicle is the induced fake objects instead of the attacker herself~\cite{shin2017illusion}.

To address such limitations, synchronized spoofing~\cite{shin2017illusion} is proposed to use the \textit{synchronization} process described above and also illustrated in Fig.~\ref{fig:sync_and_async_attack}.
The early-stage attack designs can inject only 10 spoofed points~\cite{shin2017illusion}, but the following works progressively increase the number of spoofed points to 60~\cite{cao2019adversarial} and 200~\cite{jiachen2020towards}, and show that the 60 and 200 spoofed points are enough to cause fake object injection on latest object detectors. After the success, the attack capability of injecting 200 points under the Chosen-Pattern Injection (CPI) assumption (detailed later in \S\ref{sec:cpi}) becomes the \textit{de facto} threat model in the following works~\cite{zhongyuan2021object, hallyburton2022security, hau2021shadow}. However, none of them have (1) demonstrated and quantified the CPI attack capability, and (2) studied their attack effect on new-gen LiDARs, which have new features such as laser timing randomization and pulse fingerprinting that can directly challenge their basic design assumption --- the requirement of synchronization. In this paper, we fill both of these critical research gaps.

\textbf{Spoofing Attacks for Object Removal.} 
With the success to achieve object injection, more recent works start to explore using LiDAR spoofing for object removal. Specifically, saturating attack~\cite{shin2017illusion} is the first to show the object removal effect. In this attack, instead of firing pulsed lasers like the prior works above, it fires a strong \textit{continuous} laser to the victim LiDAR to indirectly cause measurement errors of laser-receiving events. However, due to the requirement of maintaining continuous high-power laser, it is physically difficult for the attack laser beam to (1) achieve a large receiver area coverage at the victim LiDAR side with high intensity, and (2) maintain the attack effect for a long time. For example, the demonstrated object removal effect is only about removing a 41$\times$42 cm$^2$ metal plate, lasting $<$4 seconds, which thus makes it highly limited in real-world attack scenarios such as when attacking AD. 
\newpart{For example, the saturating attack (0.8 W average power) needs roughly 10 times higher average power than the HFR attack (80 W peak power) (\S\ref{sec:high_freq_attack}) to achieve similar attack effectiveness, but such a powerful diode is not publically available so far.}

To address such limitations, similar to the object injection side, more recent works start to leverage synchronized spoofing techniques. Specifically, the ORA attack (object removal attack)~\cite{zhongyuan2021object} found that using the synchronized spoofing capability of injecting 200 points under the CPI assumption as mentioned above, attackers can fool 3D object detectors by strategically injecting spoofed points inside the target object's bounding box, since the point cloud of legitimate objects should have points mostly on the object surface instead of inside. However, such removal effect still depends on the object detector-side vulnerabilities. Most recently, the physical removal attack (PRA)~\cite{cao2023you} found that the object removal effect can be more generally achieved by moving all points on a victim object to within the minimum operational threshold (MOT) of the victim LiDAR, which is common filtering mechanism to automatically discard points below a certain distance. This filtering is implemented in most LiDARs because the LiDAR measurement at a very close distance is generally inaccurate since the ToF can be too short to distinguish from errors due to inadequate calibration. This work shows the capability of removing $\sim$4k points and thus can remove critical road objects such as pedestrians, which is thus the current state-of-the-art in object removal attack capability.

However, similar to the object injection attack side, none of these prior works have considered the new-gen LiDARs that may come with features that can break some of their fundamental design assumptions. In this paper, we find that (1) PRA can no longer be applied to new-gen LiDARs due to the requirement of synchronization (\S\ref{sec:lidar_removal}), and (2) the assumption of non-zero MOT for PRA may not necessarily hold (e.g., for XT32~\cite{XT32}, \S\ref{sec:specific-lidars}). We further identify a new object removal attack adapted from the saturation attack, which does not require synchronization but can still achieve practical object removal effects (can remove $>$5k points in 10$\times$10 m$^2$ area, which is enough for 5 cars as shown later in~\S\ref{sec:od_removal}).

\label{sec:async_attack}

\label{sec:sync_injection_attack}

\label{sec:sync_removal_attack}

\nsubsubsection{Chosen-Pattern Injection (CPI) Attack Capability}
\label{sec:cpi}
So far, all prior works using synchronized spoofing for object injection explicitly or implicitly assume that at the actual attack time the attacker is able to accurately inject a specific spoofed point cloud pattern chosen by the attacker beforehand, e.g., via various kinds of off-line optimization/identification processes~\cite{cao2019adversarial, jiachen2020towards, hallyburton2022security}. In this paper, we thus explicitly define this as the \textit{Chosen Pattern Injection (CPI)} attack capability.

Such an attack capability is theoretically achievable since the synchronized LiDAR spoofing technique should by design be capable of precisely controlling the position of each spoofed point. However, so far no prior works have systematically studied how achievable this is in practice, not to mention to provide a systematic quantification of the pattern control capability, which is necessary for achieving valid security analysis on the object detector side. In this paper, we thus perform the first measurement study to fill this critical gap, with the coverage of both first- and new-gen LiDARs.

\nsubsection{Threat Model} \label{sec:threat_model}

We follow the same threat model as in prior works~\cite{cao2019adversarial, jiachen2020towards, hallyburton2022security}, i.e., the attacker fires malicious lasers from their spoofer to the victim LiDAR (\S\ref{sec:lidar_spoofing_attack}, Fig.~\ref{fig:sync_and_async_attack}).
As described in~\cite{hallyburton2022security}, the spoofer device can be at a front vehicle, vehicle in the next lane, or a roadside in AD scenarios.

\nsection{Measurement Study Setup
} \label{sec:methodology}
\vspace{0.1in}

\nsubsection{Targeted LiDARs}
Table~\ref{tbl:target_lidars} summarizes the LiDARs involved in this measurement study. As shown, we are able to cover 9 popular LiDARs in total, including 3 first-gen ones, 6 new-gen ones, an operating range from 9 m to 300 m, 3 scanning types (Rotating, MEMS, and Flash), and 3 security-related features (simultaneous laser firing, timing randomization, and fingerprinting). This makes this work the largest-scale measurement study on this topic so far in terms of both LiDAR numbers (none of the prior works studied over 1 LiDAR model, while we study 9) and LiDAR types (all prior works concentrate on first-gen ones, and we are the first to study the new-gen ones).

\nsubsection{Targeted Object Detection Models and Datasets}
\label{sec:target_lidar_od}
Our study includes 5 popular DNN-based 3D object detectors in total covering all the 3 major model types (\S\ref{sec:3d_obj}): voxel-based (PointPillars~\cite{lang2019pointpillars}, SECOND~\cite{yan2018second}, and Part-A$^2$~\cite{shi2020points}), point-based (3DSSD~\cite{yang20203dssd}), and point voxel-based (PV-RCNN~\cite{shi2020pv}). In order to study the impact of the training dataset, we also prepare 5 different versions of PointPillars~\cite{lang2019pointpillars} trained on 5 different datasets (KITTI~\cite{Geiger2012CVPR}, Waymo~\cite{Sun_2020_CVPR}, nuScenes~\cite{nuscenes}, Lyft~\cite{lyft} datasets), and the dataset used in Apollo 6.0~\cite{apollo}. This thus makes our measurement also the largest in terms of the coverage of object detection model architectures and training datasets. We directly use pre-training ones from MMdetection3D~\cite{mmdet3d2020}, OpenPCDet~\cite{openpcdet2020}, or the Apollo repository~\cite{apollo}.

\begin{table*}[t!]
\centering
\footnotesize
\caption{The 9 LiDARs involved in our measurement study. 1st-G and New-G: First- and new-generation. FOV: Field of View.}
\label{tbl:target_lidars}
\setlength{\tabcolsep}{1.2pt}
\renewcommand{\arraystretch}{0.75}
\begin{tabular}{llccccccccc}
\toprule
                  &                     & \multicolumn{3}{c}{Velodyne}         & Leddar & Ouster & Intel & Livox & Hesai & Robosense \\ \cline{3-11} 
                  &                     & \includegraphics[width=0.3in]{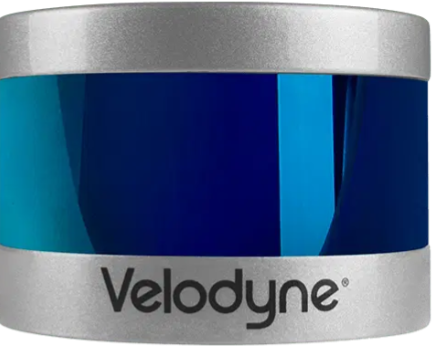}           & \includegraphics[width=0.25in]{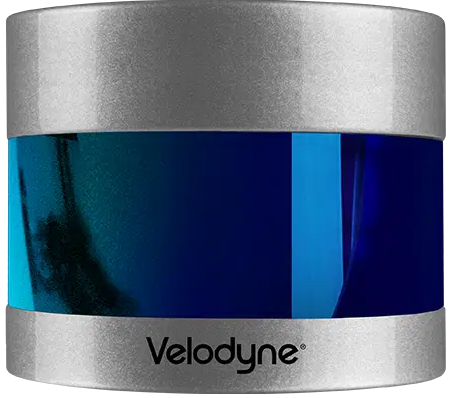}           
                  & \includegraphics[width=0.25in]{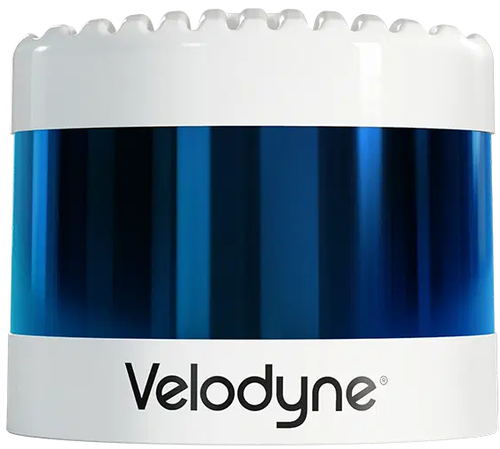}     & \includegraphics[width=0.25in]{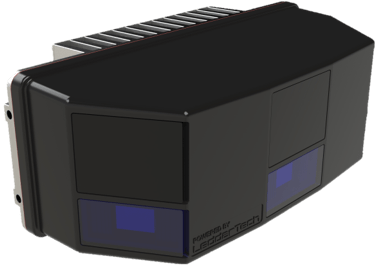}
                  & \includegraphics[width=0.2in]{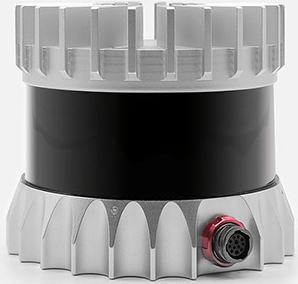}      & \includegraphics[width=0.2in]{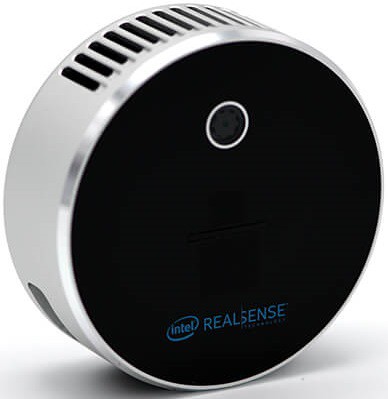}
                  & \includegraphics[width=0.25in]{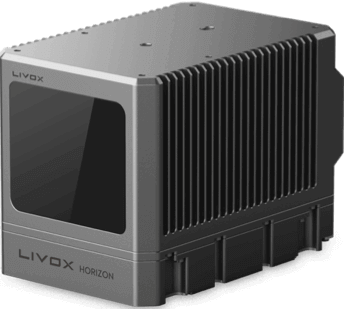}    & \includegraphics[width=0.25in]{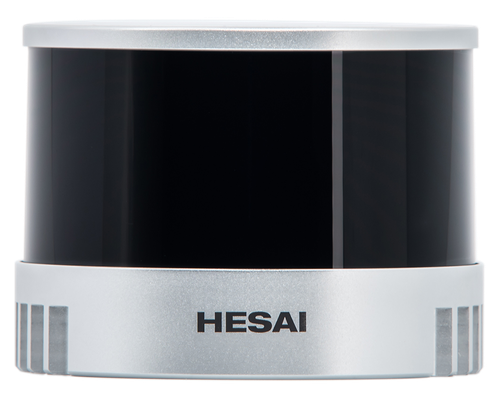}       
                  & \includegraphics[width=0.2in]{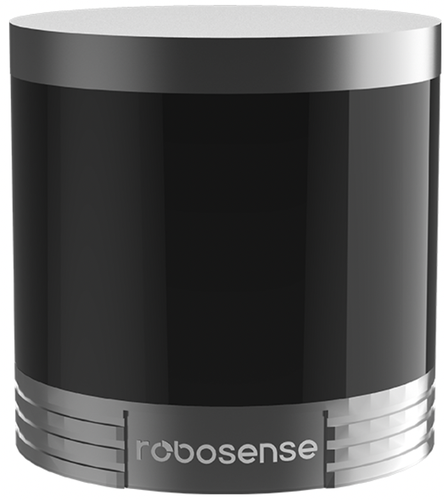}
                  \\
                  & & VLP-16~\cite{VLP16}     & VLP-32c~\cite{VLP32c}    & VLS-128~\cite{VLS128}    & Pixell~\cite{pixell} & OS1-32~\cite{OS1-32} & Realsense L515~\cite{L515} & Horizon~\cite{livox_horizon} & XT32~\cite{XT32}  &  Helios 5515~\cite{Helios}     \\ \hline
\multirow{10}{*}{\vspace{0.15in}\rotatebox{90}{\notsotiny{General Specs}}} & Gen. (year)       & 1st-G \notsotiny{(2016)}   &  1st-G \notsotiny{(2017)}   & 1st-G \notsotiny{(2017)}   & New-G \notsotiny{(2019)} & New-G \notsotiny{(2019)} & New-G \notsotiny{(2019)} & New-G \notsotiny{(2020)}  &  New-G \notsotiny{(2020)}  & New-G \notsotiny{(2021)}               \\
                  & Scanning Type       & Rotating      & Rotating      & Rotating      & Flash         & Rotating       & MEMS          & MEMS           & Rotating      & Rotating      \\
                  & Wavelength          & 905 nm        & 905 nm        & 905 nm        & 905 nm        & 865 nm         & 860 nm        & 905 nm         &  905 nm       & 905 nm        \\
                  & Vertical FOV        & 30$^{\circ}$  & 40$^{\circ}$  & 40$^{\circ}$  & 16$^{\circ}$  & 45$^{\circ}$   & 55$^{\circ}$  & 25.1$^{\circ}$ & 31$^{\circ}$  & 70$^{\circ}$  \\
                  & Horizontal FOV      & 360$^{\circ}$ & 360$^{\circ}$ & 360$^{\circ}$ & 180$^{\circ}$ & 360$^{\circ}$  & 70$^{\circ}$  & 81.7$^{\circ}$ & 360$^{\circ}$ & 360$^{\circ}$ \\
                  & Max. Range {[}m{]}  & 100           & 200           & 300           & 56            & 120            & 9             & 260            & 120           & 150           \\
                  & Min. Range {[}m{]}  & 1             & 1             & 0.5           & 0.1           & 0.3            & 0.25          & 0.5            & 0             & 0.2           \\
                  & Vertical Channel    & 16            & 32            & 128           & 8             & 32             & -             & -              & 32            & 32            \\
                  \hline
\multirow{3}{*}{\vspace{-0.08in}\rotatebox[origin=c]{90}{
 \notsotiny{Security}
}}                & Simul. Firing       & 1             & 2             & 8             & 3             & 32             & 1             & 1              & 1             & 1             \\
                  & Timing Random.      &               &               &               & \CheckmarkBold& \CheckmarkBold & \CheckmarkBold& \CheckmarkBold &               & \CheckmarkBold\\
                  & Fingerprinting      &               &               &               &               &                &               &                & \CheckmarkBold&               \\ \toprule
\end{tabular}
\end{table*}

\nsubsection{Targeted LiDAR Spoofing Attacks}\label{sec:target_attacks}

Our study targets spoofing attacks with the latest attack capabilities on both the object injection and removal sides (\S\ref{sec:attack_taxonomy}). Specifically, for object injection, we reproduce the latest synchronized spoofing technique~\cite{cao2019adversarial,jiachen2020towards,cao2023you}, and for object removal, we reproduce the latest physical removal attack (PRA)~\cite{cao2023you}. In addition, since the latest removal attack cannot be applied when synchronization is not possible (which we found to be the case for new-gen LiDARs in general, detailed in~\S\ref{sec:inj_other_lidars}), we further identify and include in our measurement a new asynchronized spoofing technique for object removal, which is adapted from the saturation attack (\S\ref{sec:attack_taxonomy}) but with much more practical object removal capabilities:

\textbf{High-Frequency Removal (HFR): New Asynchronized Spoofing Technique for Object Removal.} \label{sec:high_freq_attack}
As described in~\S\ref{sec:attack_taxonomy}, due to the requirement of maintaining continuous high-power laser, the saturation attack is fundamentally limited in the removal area size and duration (e.g., 41$\times$42 cm$^2$ in $<$4 seconds~\cite{shin2017illusion}), making it highly limited in real-world attack scenarios. Later works address this by using synchronization, but this also makes them inapplicable to new-gen LiDARs (\S\ref{sec:inj_other_lidars}). To address this, we identify a new adaptation of the saturation attack, which still does not require synchronization, but instead of using high-power continuous laser, it uses \textit{high-frequency pulsed laser}. This new attack is thus called \textit{high-frequency removal (HFR) attack}. The key idea is to fire a large number of attack laser pulses to the victim LiDAR at a frequency that is higher than the laser-firing frequency of the victim LiDAR. This allows the attack laser to hit every laser-firing event of the victim LiDAR in the scanning range hit by the spoofer, which can thus achieve the spoofing effect for every points in that range without any knowledge of the victim scanning pattern (i.e., without requiring synchronization). Fig.~\ref{fig:sync_and_async_attack} illustrates the spoofing mechanism, and Fig.~\ref{fig:highfreq_vs_saturating} in Appendix illustrates the difference to the saturation attack. Due to the lack of synchronization, the receiving timing of the attack laser is random, and thus the spoofing effect will be moving each legitimate surface point of the targeted objects to a random position or undetectable area of the victim LiDAR (e.g., within MOT). As shown later in~\S\ref{sec:od_removal}, this can completely destruct the point cloud patterns at the original object position and thus cause the object removal effects.

\nsubsection{LiDAR Spoofer Setup}
\label{sec:optical_setup}
\label{sec:arbitrary_point_injection}
\label{sec:spoofer_design}

\begin{figure}[t!]
\centering
\includegraphics[width=\linewidth]{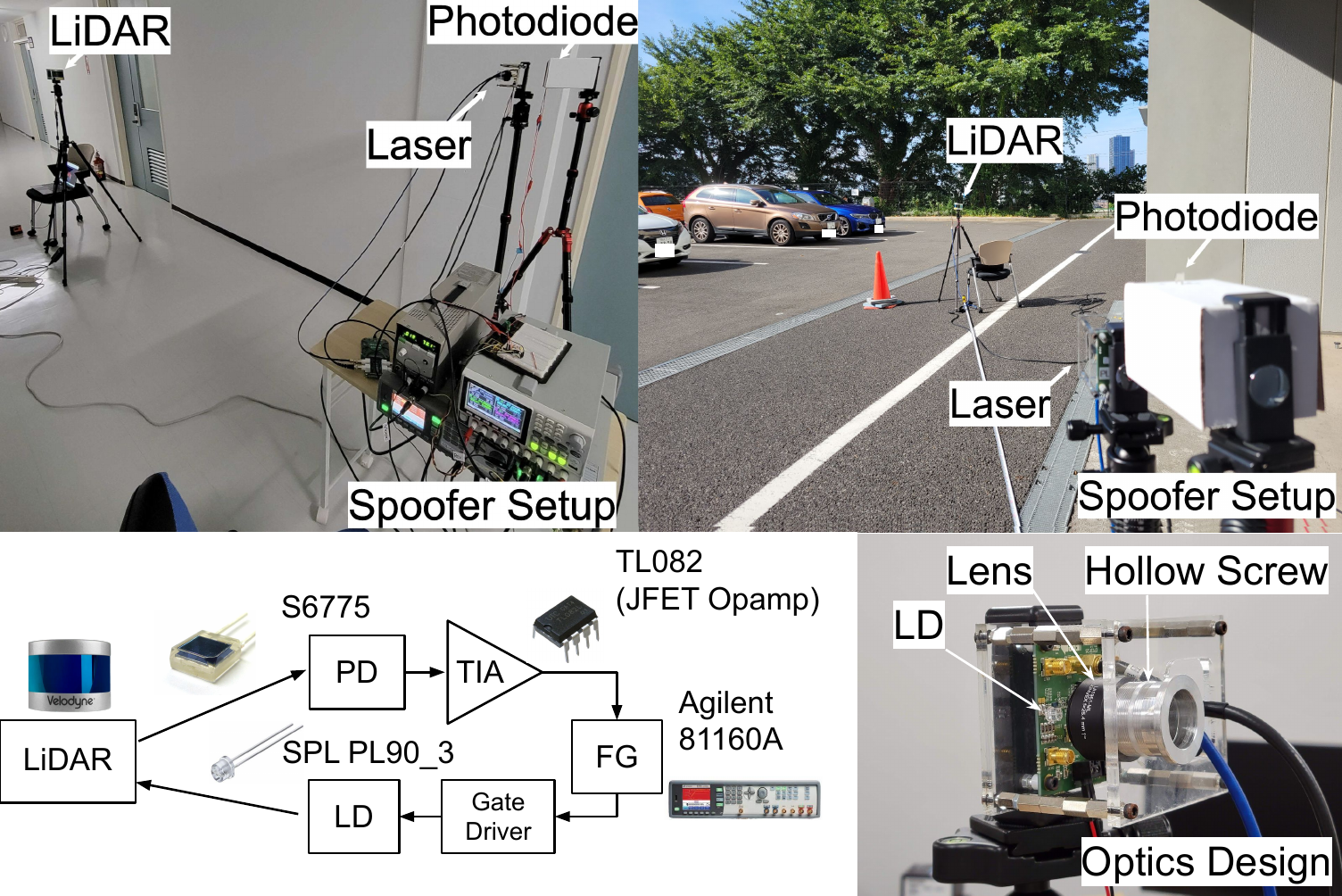}
\caption{Overview of our LiDAR spoofer setup, the optics design, and the setup of the indoor and outdoor experiments. PD: Photodetector. TIA: Transimpedance amplifier. FG: Function generator. LD: Laser diode. More details in Appendix~\ref{appendix:spoofer}.
}
\label{fig:spoofer1}
\end{figure}

Fig.~\ref{fig:spoofer1} shows an overview of the LiDAR spoofer, its optics design, and the setups of the indoor and outdoor experiments to support our measurement study of the targeted spoofing attacks (\S\ref{sec:target_attacks}). We generally follow the setup adopted in the latest works~\cite{cao2019adversarial, jiachen2020towards, hallyburton2022security, cao2023you}, so we leave the details to Appendix~\ref{appendix:spoofer}. 

\textbf{Spoofer Improvements.} When reproducing the prior works, we also implemented several improvements on top of the latest spoofer setup to more properly study the spoofing attack capabilities. For instance, we find that inadequate optical design can significantly affect the number and angle coverage of spoofable points due to undesired diffusion and convergence of the attack laser beam. Specifically, if the laser beam is expanding, the intensity rapidly decays with distance, which thus limits the number of spoofable points; if the laser beam is converging, the diameter of the beam decreases with the distance, which thus makes it difficult to aim and cover a wide receiver angle. Ideally, the laser beam should be collimated without diffusion and convergence to achieve the target LiDAR with a minimum loss. 
To achieve this, we use a 25.4 mm focal length plano-convex lens with 1 inch (25.4 mm) diameter to cover the laser emitted by SPL90\_3~\cite{SPLPL90_3}, which has a maximum beam divergence angle of 25$^{\circ}$. 
Since the diameter of the laser at the focal point is $\tan{25^\circ}\times 25.4 \times 2 = 23.7$ mm, we can cover it with the 1 inch (25.4 mm) lens. 
To precisely calibrate the lens setup, we develop a device that can use a hollow screw to adjust the distance between the laser diode (LD in Fig.~\ref{fig:spoofer1}) and the lens.
Technically, this allows the spoofer to maintain a stably large number and angle coverage of spoofable points from hundreds of meters away. 

Besides the optics side, we also notice that inadequate electronic implementations in the spoofer setup can introduce inaccurate laser detection timing (from LiDAR to PD in Fig.~\ref{fig:spoofer1}) and also long and unstable delays in the signal processing of the spoofer devices, which can thus significantly affect the attack capability in precisely controlling the placement of a spoofed point at a chosen 3D position. To address this, we improve the amplifier for the PD to increase the laser detection accuracy, and also improve the function generator (FG) setup to allow more precise nanosecond-level configuration and calibration. As shown in Fig.~\ref{fig:arbitrary_point} and detailed later in~\S\ref{sec:inj_vlp16}, these improvements significantly increase the latest spoofing attack capabilities in terms of the spoofed point number, angle coverage, pattern control, and robustness over distance. Following the best practices advocated in~\cite{shen2022sok}, we release the raw hardware design file (CAD file) and the bill of materials on our website~\cite{project_page} to make it easier for future works to reproduce and build upon (most recent works did not do this~\cite{cao2019adversarial,jiachen2020towards,cao2023you}).

\nsection{Object Injection Attack Measurements} \label{sec:injection_attack}

With the measurement setup above, in this paper we perform the first large-scale measurement study for both classes of state-of-the-art LiDAR spoofing attacks: object injection attack and object removal attack, which will be the focus of this and the next section respectively. For each attack class, we first perform the attack capability measurements at the LiDAR point cloud level, and then model such attack capabilities for the subsequent measurements at the object detector level.

\nsubsection{LiDAR-Level Measurements (RQ1, RQ2)}
\label{sec:lidar_injection}

In this section, we measure the object injection attack capabilities at the LiDAR point cloud level, which is measured by the spoofing attack's capability in injecting spoofed points. We perform this measurement using an attack laser with sufficiently high intensity so that the attack-induced spoofed points can be easily differentiated from the benign ones. To answer RQ1, we first re-visit the existing state-of-the-art spoofing attacks for point injection on the VLP-16 LiDAR~\cite{VLP16}, which is predominantly used as the \textit{only} evaluation target in prior works~\cite{shin2017illusion, cao2019adversarial, jiachen2020towards, cao2023you, hallyburton2022security}. After that, we then conduct the measurement study on new-gen LiDARs to answer RQ2.

\nsubsubsection{Re-Visiting Design Assumptions Made in Prior Works with VLP-16 (RQ1)} \label{sec:inj_vlp16}

Prior works have shown that the number of spoofed points in VLP-16 increases as the quality of the spoofer device improves, ranging from 60~\cite{cao2019adversarial} to $\sim$4k~\cite{cao2023you}. 
As shown in Table~\ref{tbl:distance}, with our spoofer improvements (\S\ref{sec:spoofer_design}), such attack capabilities are further improved significantly,
which can now inject $>$6,131 points indoors and $>$6,514 points outdoors. This is at least 50\% more than those in the latest prior work~\cite{cao2023you}, which capped at $\sim$4,000. We noticed a concurrent work that may have also made contributions in improving the spoofing capability~\cite{jin2022pla}. The full paper is not available at the time of our writing, but from the abstract it seems that (1) the spoofing capability is still capped at 4200 points, which is still at least 50\% fewer than ours; and (2) the study scale is much smaller than ours in terms of LiDAR number (only 2, versus 9 for us) and likely also LiDAR types and generation coverage.

Meanwhile, we also observe that the number of spoofed points and angles in the outdoor setup is generally larger than those in the indoor setup. We consider this reasonable since the legitimate laser reflections are fewer in the outdoor environment (e.g., no wall reflections), leaving more room for spoofing.  However, this actually contradicts those reported results in the latest prior work~\cite{cao2023you}: as shown in Table~\ref{tbl:distance}, the spoofed points outdoors are significantly fewer than those indoors (1.8k versus 4k) and they attribute this to the lighting condition differences (e.g., they also report that the number of spoofed points decreases from $\sim$2,500 at night to $\sim$1,600 at day.)
We consider that our results are achieved by our more careful optical setup as described in~\S\ref{sec:optical_setup} since the spoofing laser intensity is much stronger than the sunlight and should not be decayed in a short 10 m flight. 
We suspect that the optical setup of the previous work is not well calibrated and the laser beam diverges.

\begin{observation}{RQ1}
With electronic and optical setups, the LiDAR spoofing attack can inject $>$6,000 points ranging $>$80$^{\circ}$, which is a significantly higher number of points and wider range than in previous studies.
Furthermore, previous observations that spoofing distance and lighting conditions can greatly affect the LiDAR spoofing capability~\cite{cao2023you} no longer hold if implemented using more careful optics.
\end{observation}

\nsubsubsection{CPI Attack Capability Measurements} \label{sec:cpi_capability}
As shown in Fig.~\ref{fig:arbitrary_point}, with our spoofing improvements we are able to demonstrate strong spoofed point pattern control capability for the first time. Considering that the CPI attack capability is a common design-level assumption made in prior works (\S\ref{sec:cpi}), we thus perform the first systematic quantification of it to allow more rigorous attack capability analysis on the object detector side.
Note that some prior works tried to include a modeling of such CPI attack capability in their object detector-side analysis~\cite{hallyburton2022security}, but their modeling is based on their intuitions (e.g., simply add vertical and horizontal noises~\cite{hallyburton2022security}) instead of the LiDAR sensing mechanisms and the spoofer designs. As shown later in~\S\ref{sec:od_injection_eval}, such erroneous spoofing accuracy modeling can cause significant differences in the object detector-side attack results.

\begin{figure}[t!]
\centering
\vspace{-0.1in}
\includegraphics[width=\linewidth]{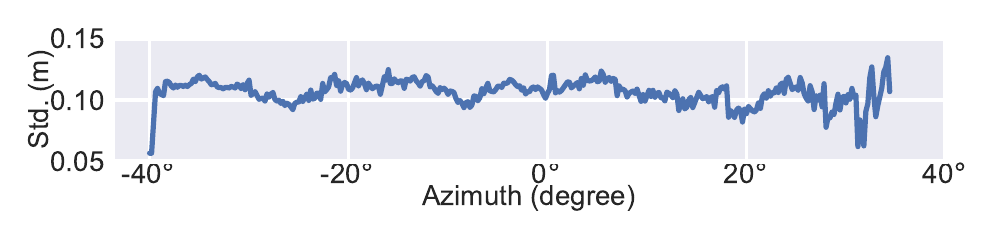}
\vspace{-0.3in}
\caption{Standard deviations of inner-frame error on VLP-16.}
\label{fig:inner_frame_error}
\end{figure}

\begin{figure}[t!]
    \begin{minipage}{.63\linewidth}
\centering
\includegraphics[width=\linewidth]{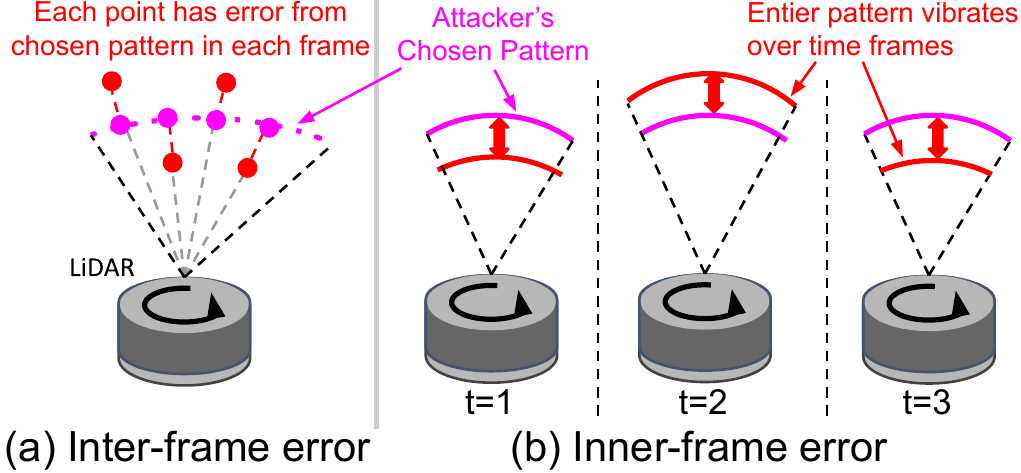}
\vspace{-0.1in}
\caption{Illustration of inner- and inter-frame errors. Inner-frame error causes spoofing inaccuracy along with the ray direction within a frame. Inter-frame error causes the entire pattern to vibrate across consecutive frames.
}
\label{fig:inter_inner}
    \end{minipage}
    \hspace{0.01in}
    \begin{minipage}{.33\linewidth}
        \vspace{-0.1in}
        \centering
        \includegraphics[width=\linewidth]{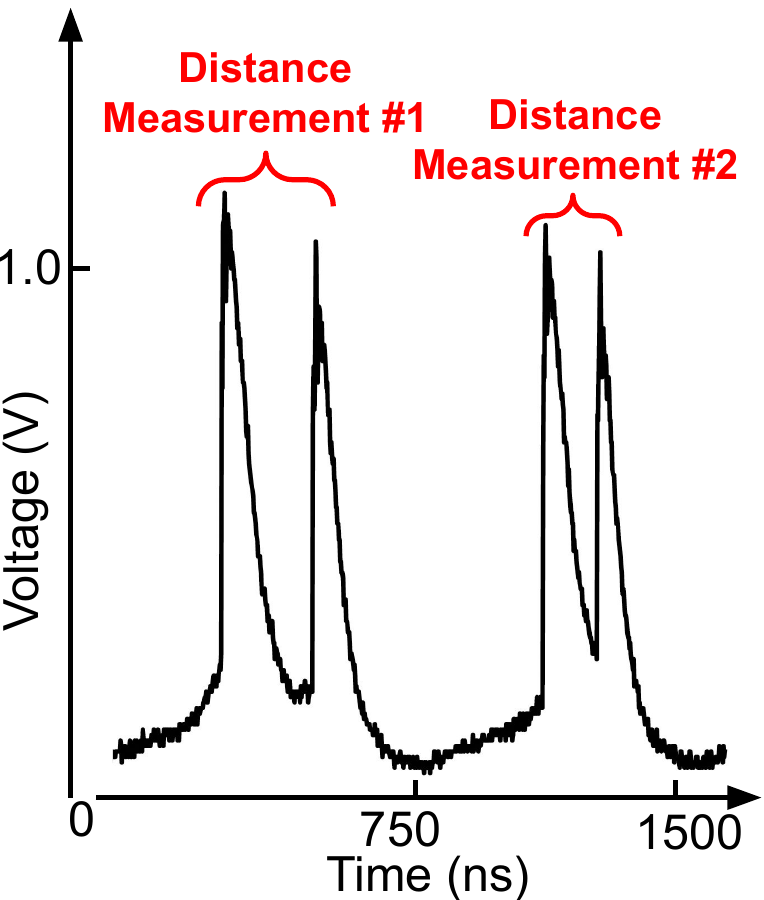}
        \caption{Examples of the receiving pulse shape of XT32~\cite{XT32}.
        }
        \label{fig:hesai_fingerprint}
   \end{minipage}
\end{figure}

Based on our attack reproduction and spoofing inaccuracy cause investigations, we find that there are two types of errors in controlling the position control of each spoofed point $x_{ij}$ at $i$-th altitude and $j$-th azimuth (illustrated in Fig~\ref{fig:inter_inner}): (1) \textit{inner-frame error} $\delta^{\rm inner}_{ij}$: the inaccuracy in spoofing a point at a chosen 3D position within an individual frame; and (2) \textit{inter-frame error} $\delta^{\rm inter}$: the inaccuracy in maintaining the spoofed position of the same point in the chosen pattern across consecutive frames. Rooted in the spoofing design (Fig.~\ref{fig:spoofer1}), the inner-frame error is due to the inaccurate signal bursting of FG to the gate driver, while the inter-frame one is due to inaccurate timing of triggering the FG from TIA. With our experimental setup, we measure both and find that $\delta^{\rm inner}_{ij}$ and $\delta^{\rm inter}$ are $\sim$10 cm as in Fig~\ref{fig:inner_frame_error} and $\sim$35 cm respectively.

\begin{observation}{RQ1}
The CPI attack capability, which is commonly assumed but never demonstrated in prior works~\cite{cao2019adversarial, jiachen2020towards, cao2023you, hallyburton2022security, zhongyuan2021object}, can be achievable in practice with a well-calibrated spoofer, while by design such pattern control capability is subject to inner- and inter-frame errors.
\end{observation}

\textbf{Assumption that object detector-level vulnerability exploitation is needed:} Combining these spoofing capability advances above (significantly improved spoofable points, angle, and CPI attack capability), this actually means that \textit{an attacker does not always need to exploit object detector-level vulnerabilities to achieve a near-front road object injection}, which is actually the major design assumption for all attack works so far on the object injection attack side~\cite{cao2019adversarial, jiachen2020towards, hallyburton2022security}.
Specifically, as clearly stated in prior works~\cite{cao2019adversarial, jiachen2020towards}, a valid point cloud for a front-near vehicle needs $\sim$2,000 points and 15$^{\circ}$ azimuth ($\theta$) coverage, but their spoofers can only achieve $<$200 points and $<$8$^{\circ}$, which is why object detector-level vulnerability exploitation (and thus those novel spoofing pattern optimization/identification) are needed. \newpart{These results indicate that the exploitation of the object detector-level vulnerability is not the necessary condition although it may help to design more sophisticated attacks.}

\begin{observation}{RQ1}
\newpart{
Now with the CPI attack capability demonstrated with sufficiently larger spoofable point number and angle, an attacker actually has sufficient capabilities to directly inject a near-front vehicle pattern. This finding has significant implications for the current line of research, since this  (1) may not need to exploit object detector-level vulnerabilities on the attack sides~\cite{cao2019adversarial, jiachen2020towards, hallyburton2022security}; and 
(2) may suggest that model-level defenses that try to leverage the inability of spoofers in directly injecting indistinguishable patterns as benign cases (e.g., \cite{jiachen2020towards}) can be ineffective.
}
\end{observation}

\begin{table}[t!]
\centering
\footnotesize
\caption{Evaluation results of the synchronized injection attack on VLP-16. 
$\mathcal{N}$: Number of injected points by spoofing. $\theta$: Azimuthal range of the point injection. $\mathcal{R}$: Point injection success rate within $\theta$.
The distance $d$ is between the spoofer and the LiDAR. Number in parenthesis: $\mathcal{N}$ from latest prior work~\cite{cao2023you}. ``-'': Data not available.
}
\label{tbl:distance}
\footnotesize
\setlength{\tabcolsep}{4.5pt}
\renewcommand{\arraystretch}{0.75}
\begin{tabular}{cccccccc}
\toprule
     & \multicolumn{3}{c}{Indoor} &  & \multicolumn{3}{c}{Outdoor (Daytime: 70 lux)} \\ \cline{2-4} \cline{6-8} 
$d$     & $\mathcal{N}$     &  $\mathcal{R}$     & $\theta$         &  & $\mathcal{N}$     &  $\mathcal{R}$     & $\theta$      \\ \cline{1-4} \cline{6-8} 
2 m  & 6,523 (-)  & 98.5\% & 82.7$^{\circ}$  &  & 7,705 (-)  &  94.9\%     & 100.5$^{\circ}$        \\
(2.5 m)  & ($<$4k)  &  &   &  & (-)  &      &        \\
4 m  & 6,386 (-)  & 96.9\% & 82.5$^{\circ}$ &  &  7,950 ({$<$1.8k})  &   96.9\%     & 101.5$^{\circ}$        \\
6 m  & 6,575 (-)  & 98.6\% & 83.4$^{\circ}$  &  & 7,357 ({$<$1.5k}) &   87.2\%    & 99.6$^{\circ}$        \\
8 m  & 6,213 (-)  & 93.8\% & 82.8$^{\circ}$  &  & 6,702 ({$<$1k}) &   97.7\%     & 83.4$^{\circ}$         \\
10 m & 6,131 (-) & 93.2\% & 82.1$^{\circ}$  &  & 6,514 ({$<$1k}) &    93.3\%     &  84.2$^{\circ}$     \\ \toprule
\end{tabular}
\vspace{-0in}
\end{table}

\nsubsubsection{Measurements on New-Gen LiDARs}  \label{sec:inj_other_lidars}

Table~\ref{tbl:inject_attack} shows the measurement results of the point injection capabilities on not only first-gen LiDARs such as VLP-16 but also new-gen ones. The first thing we find is that the latest spoofing attack for object injection, synchronized spoofing  (\S\ref{sec:attack_taxonomy}), can only be applied to the first-gen ones; for the new-gen ones, 
since they all have either timing randomization or pulse fingerprinting, synchronization (and thus CPI) becomes virtually impossible since timing randomization by design prevents the prediction of the future laser firing timing from the victim LiDAR, while pulse fingerprinting by design prevents the prediction of the future laser pulse pattern of each measurement. To still measure their vulnerabilities to spoofing attacks, we try to inject as many spoofed points as possible with a random attack with high-frequency pulses. 
The random attack is similar to the newly-identified HFR attack (\S\ref{sec:high_freq_attack}), but the frequency is tuned to achieve the largest number of injected points.

As shown, compared to the first-gen LiDARs, the security-related features in the new-gen LiDARs result in huge differences in the point injection capabilities: $>$19k injected points for Horizon, $\sim$3.2k for Helios, 100-350 on Realsense L515 and XT32, and only 28 on OS1-32. We will closely investigate such differences in the next section. Despite these variances, one observation is in common: VLP-16 is actually the \textit{only} LiDAR among all these 7 ones that can achieve the CPI attack capability --- it is not possible by design for the 6 new-gen ones due to timing randomization and pulse fingerprinting, while it is also not achievable to the other first-gen LiDAR (VLP-32c) due to simultaneous firing (\S\ref{sec:sec_enchance_feats}).

\begin{observation}{RQ1}
VLP-16 is actually the only LiDAR for which the CPI attack capability is feasible, which is the key design assumption made in all prior works on object injection attack side~\cite{cao2019adversarial, jiachen2020towards, hallyburton2022security}. This directly challenges the validity of all these existing designs against the more general and recent set of LiDARs.
\end{observation}

\nsubsubsection{Impacts from Security-Related Features}  \label{sec:sec_enchance_feats}
~

\begin{table}[t!]
\centering
\footnotesize
\caption{Measurements of the point injection capabilities on different LiDARs. For the first-gen LiDARs, the attack is synchronized spoofing (\S\ref{sec:attack_taxonomy}). For the new-gen ones, we inject as many points as possible with random firing (1 MHz). Symbols are the same as in Table~\ref{tbl:distance}. %
}
\label{tbl:inject_attack}
\setlength{\tabcolsep}{3.1pt}
\setlength{\aboverulesep}{0pt}
\setlength{\belowrulesep}{0pt}
\renewcommand{\arraystretch}{0.9}
\begin{tabular}{c|cc|cccc|c}
\toprule
       & \multicolumn{2}{c|}{\multirow{2}{*}{First-Gen}} & \multicolumn{5}{c}{New-Gen}  \\  \cline{4-8}
       &  & & \multicolumn{4}{c|}{w/ Timing Randomization}    & w/ Fingerprint  \\  \cline{2-8}
 &       VLP-16  & VLP-32c & OS1-32 & Helios    & Horizon  & L515    & XT32  \\  
 \hline  \hline$\mathcal{N}$     & 6,523    & 9,711    & 28     & 3,203   & 19,182    & 321   & 113    \\
$\mathcal{R}$     & 98.50\% & 82.90\% & 43.80\%   & 19.4\%& 79.90\%  & 0.1\%  & 2.10\%  \\
$\mathcal{\theta}$ & 82.7$^{\circ}$ & 73.2$^{\circ}$ & 0.72$^{\circ}$  & 34.2$^{\circ}$ & 103.4$^{\circ}$ & 81.7$^{\circ}$ & 70$^{\circ}$\\ \toprule
\end{tabular}
\raggedright
\vspace{0.05in}
\end{table}

\noindent\textbf{Simultaneous Laser Firing.} \label{sec:sim_fireing}
As listed in Table~\ref{tbl:target_lidars}, many LiDARs fire multiple laser pulses simultaneously. Velodyne LiDAR is likely adopting a modular approach that doubles units when increasing altitudes. Hence, the number of simultaneous laser firings is also doubled: VLP-16~\cite{VLP16} fires 1 laser during each measurement, and VLS-32~\cite{VLP32c} fires 2 lasers simultaneously. This design makes the CPI attack infeasible because we cannot selectively return the laser to each simultaneous laser due to the large diameter of the spoofing laser. For example, on VLP-32c we can always inject spoofed points pair by pair, and the injected pair will always have the same distance to LiDAR. Although simultaneous laser firing may help attackers since it can affect more points by a single attack laser pulse, the CPI becomes no longer feasible.

\noindent\textbf{Timing Randomization.}
New-gen LiDARs typically have a feature that randomizes their laser shooting timing at each firing to mitigate interference from other LiDARs as listed in Table~\ref{tbl:target_lidars}. The timing randomization makes the CPI attack capability virtually impossible because the attacker can no longer synchronize with the laser firing pattern. However, interestingly, we find that the attacker may still be able to inject some points if the randomization is not strong enough. To quantify the impacts of the timing randomization, we measure the laser firing interval (not available from data sheets) with a PD and an oscilloscope. Table~\ref{tbl:randomization} illustrates the histogram, standard deviation, and maximum value of laser firing intervals of each LiDAR with timing randomization. 
We also categorize and fit the observed firing intervals into the uniform or Gaussian distribution based on the shape of its histogram.
We convert time difference to distance with the following formula: $\frac{\Delta t \times c}{2}$, where $\Delta t$ is the timing difference and $c$ is the speed of light. Fig.~\ref{fig:livox_spoofing} in Appendix shows the point clouds of Horizon~\cite{livox_horizon}, which can illustrate the impacts of timing randomization on spoofing capability. As shown, when under attack, the whole point cloud becomes highly randomized, and the object shapes (e.g., our lab room as shown in the benign case) completely disappeared. We further evaluate the significance of the errors on object detectors in~\S\ref{sec:impact_noise}.

\begin{table}[t!]
\centering
\footnotesize
\setlength{\tabcolsep}{2pt}
\renewcommand{\arraystretch}{0.75}
\caption{Distribution of laser firing intervals for LiDAR with timing randomization. Std. $\sigma$: Standard deviation of the error caused by the timing randomization in meters. Max. $\Delta$: Maximum error caused by the timing randomization in meters. 
}
\label{tbl:randomization}
\begin{tabular}{lccccc}
\toprule
                        & OS1-32~\cite{OS1-32}        & Horizon~\cite{livox_horizon}       &  L515~\cite{L515} & Pixell~\cite{pixell}  &  Helios~\cite{Helios}    \\ \cline{2-6} 
                        & \includegraphics[width=0.5in]{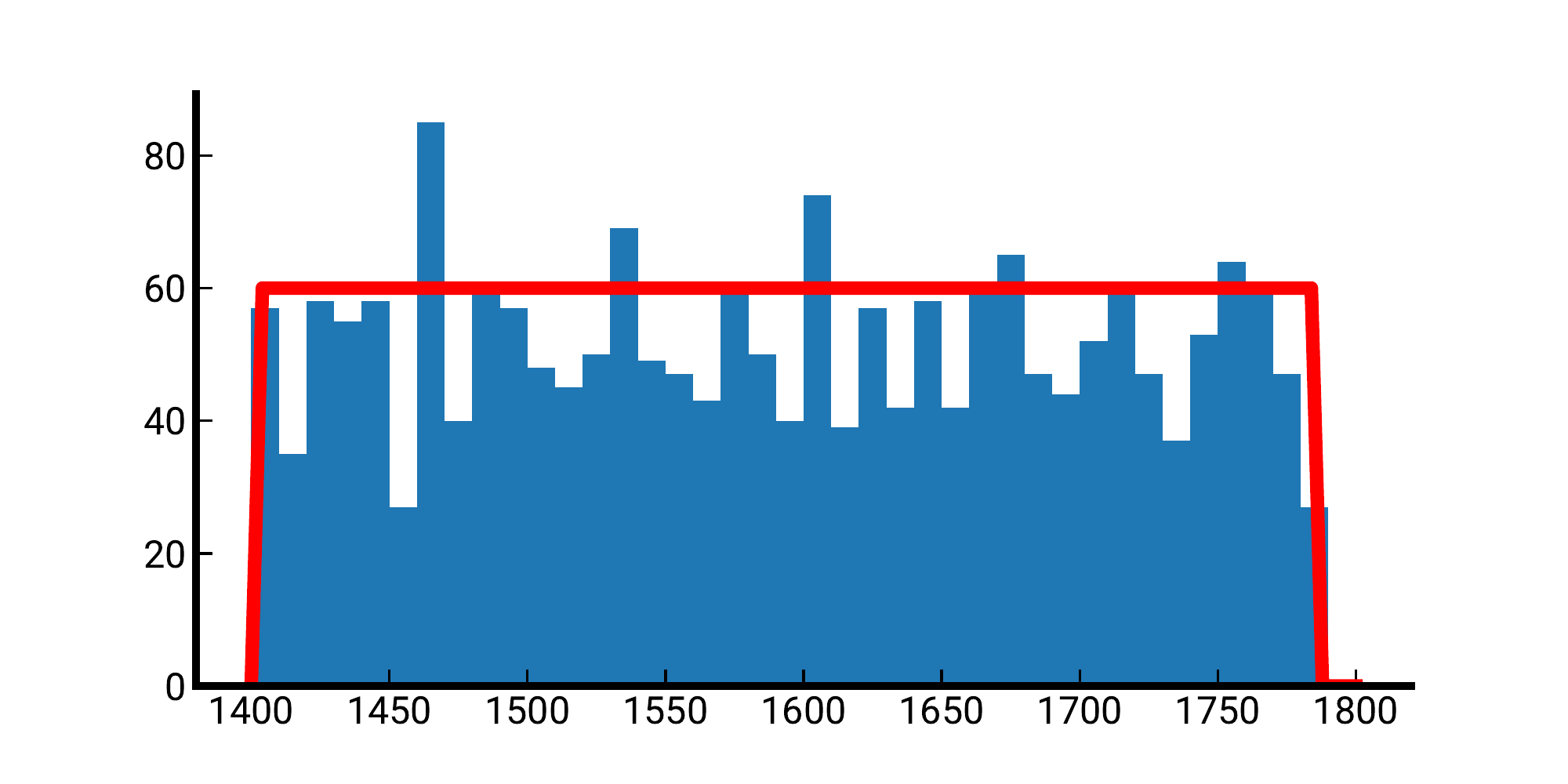}              
                        & \includegraphics[width=0.5in]{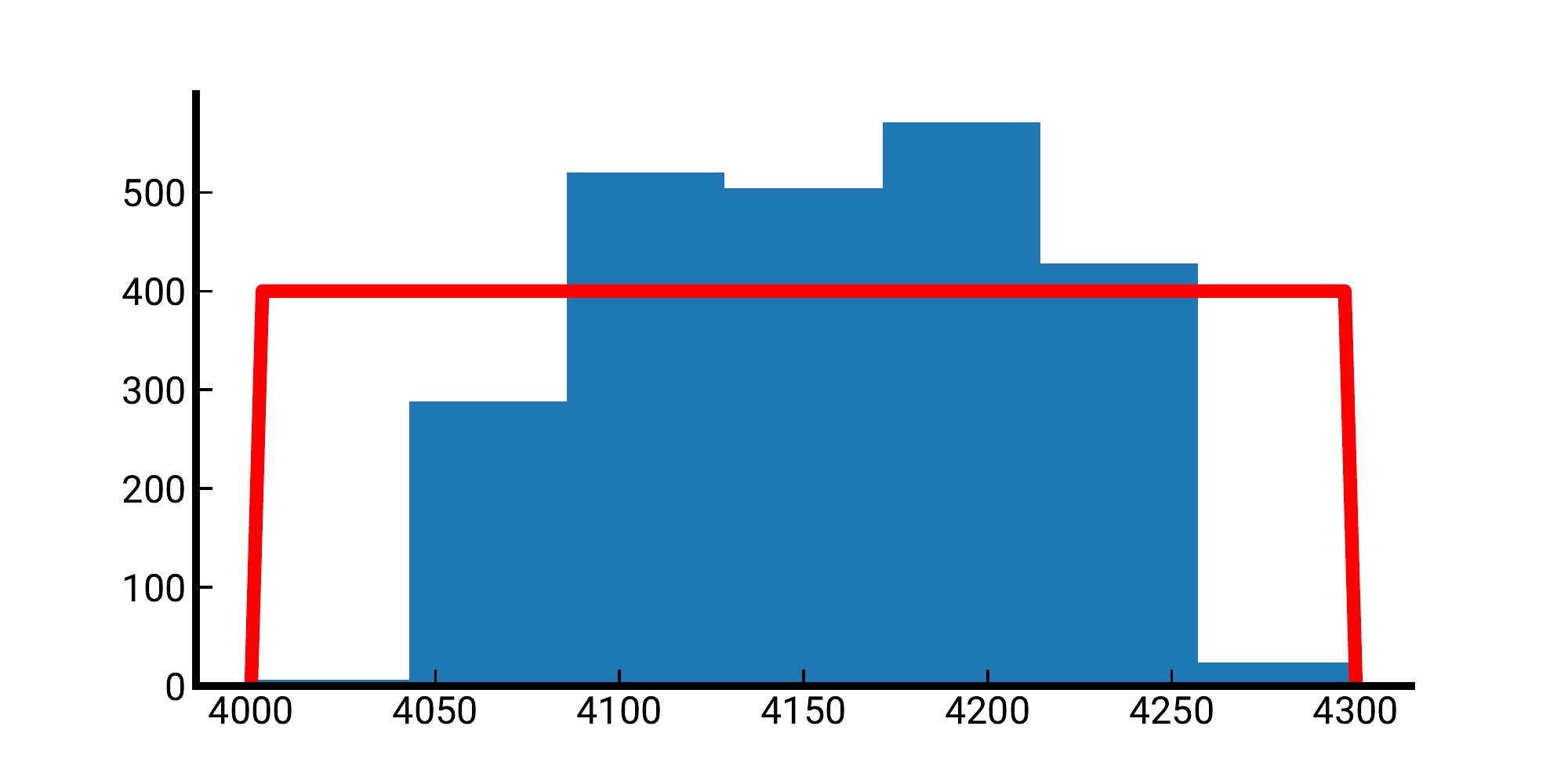}               
                        & \includegraphics[width=0.5in]{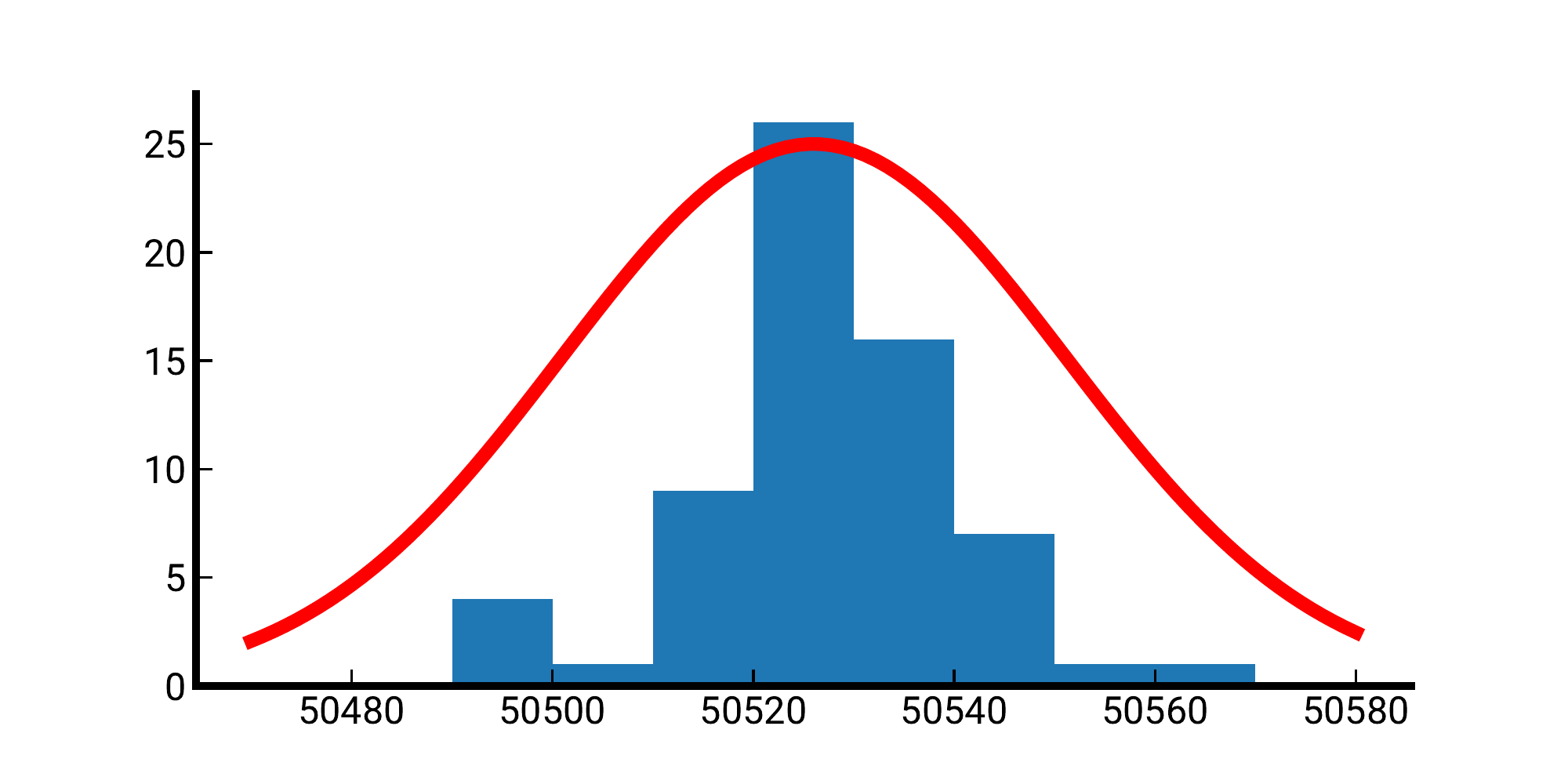}               
                        & \includegraphics[width=0.5in]{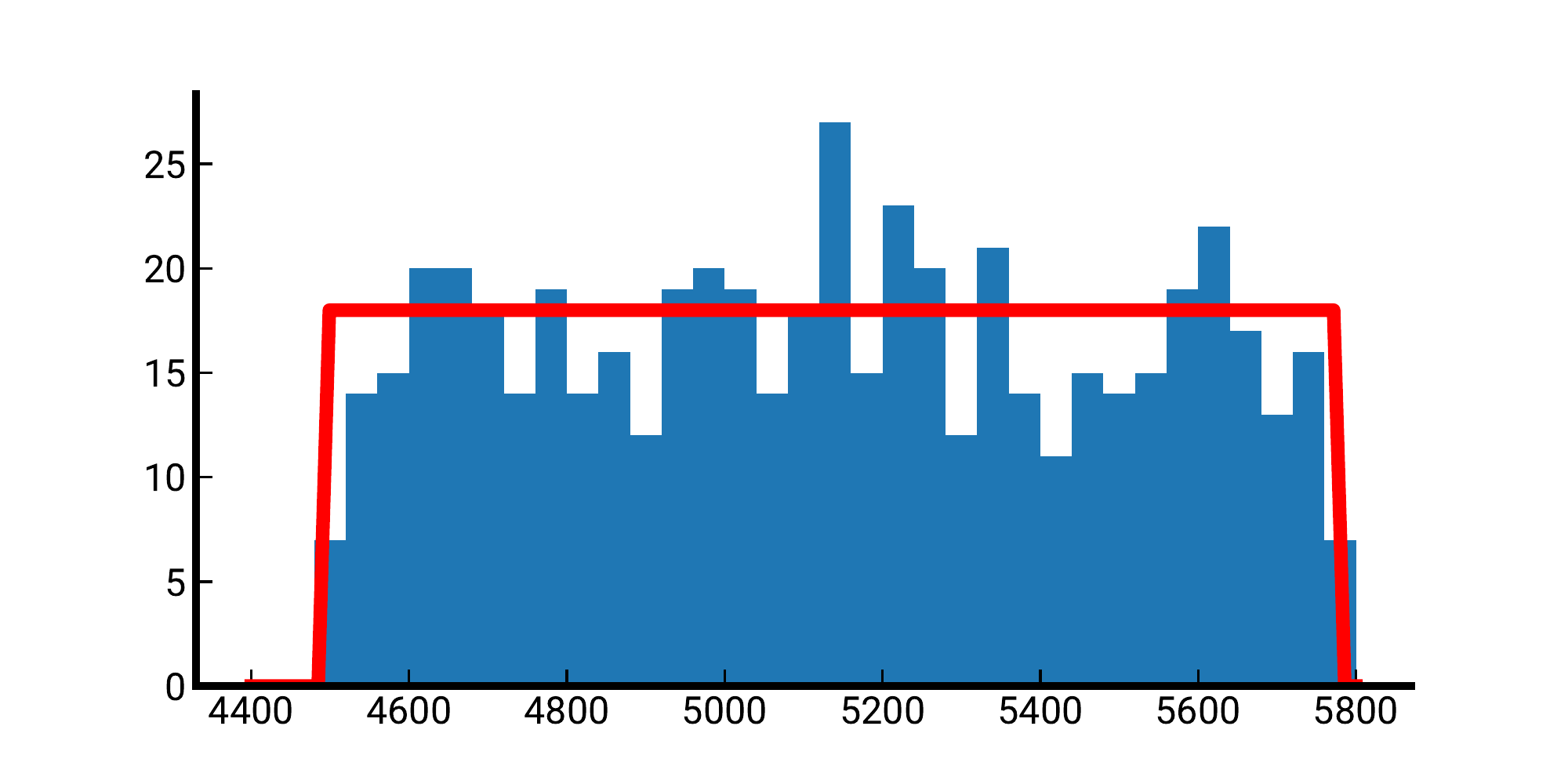}
                        &
                        \includegraphics[width=0.5in]{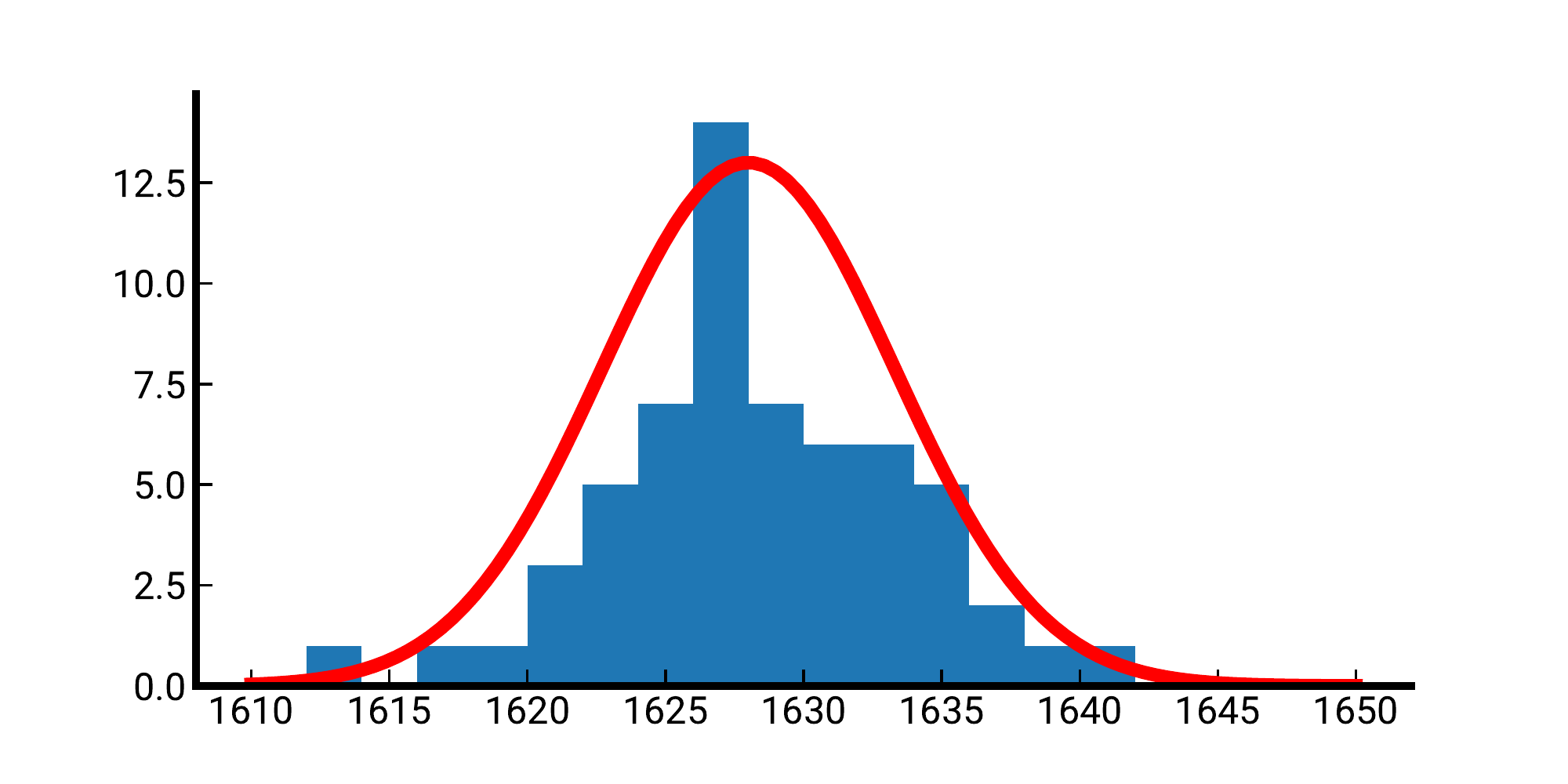}
                        \\ \hline
Dist. {[}$\mu$s{]} & $\mathcal{U}_{1.4, 1.8}$ & $\mathcal{U}_{4.0, 4.3}$ & $\mathcal{N}_{51, 0.025}$   & $\mathcal{U}_{4.5, 5.8}$ & $\mathcal{N}_{1.6, 0.005}$\\
Std. $\sigma$      & 33.3 m          & 26.0 m            & 7.5 m           & 110.4 m     &    1.5 m   \\
Max. $\Delta$      & 57.7 m          & 45.0 m            & 20.1 m           & 191.3 m     &   5.3 m      \\ \toprule
\end{tabular}
$\mathcal{U}_{{\rm min, max}}$ - Uniform distribution, $\mathcal{N}_{{\rm mean, std}}$ - Gaussian distribution
\vspace{0.1in}
\end{table}

\noindent\textbf{Pulse Fingerprinting.} We identify that Hesai XT32~\cite{XT32} can foil synchronization (and CPI attack capability) even without timing randomization due to its pulse fingerprinting. While we cannot be entirely sure due to the lack of official documentation, XT32 likely encodes its fingerprinting in the interval of the pair: As shown in Fig.~\ref{fig:hesai_fingerprint}, the XT32 pulse always forms a pair of pulses (two closely-connected spikes) corresponding to a single distance measurement.
However, we also find that the fingerprinting itself cannot perfectly defend against spoofing attacks because random spoofing can sometimes coincide with the fingerprinting interval, which can lead to up to 113 spoofed points as in Table~\ref{tbl:inject_attack}.
This could be because such pulse fingerprinting is originally developed for anti-interference purposes (e.g., to allow multiple LiDARs to operate at close range) instead of security and thus does not have enough randomness/entropy. Furthermore, it is not trivial to design more complex fingerprinting while ensuring eye safety as we will discuss in~\S\ref{sec:discussion}.
This means that if in each attack laser-firing event the attacker also fires a pair of pulses with the interval that can most likely coincide the fingerprinting interval (e.g., the one we found that can spoof 113 points), there can still be a random subset (e.g., 113) of the points in the attacker-chosen point cloud pattern that can be spoofed with the CPI attack capability. 
Thus, later in~\S\ref{sec:od_injection}, we model this effect on the spoofing capability as a random downsampling from a chosen pattern.

\begin{observation}{RQ2}
The current pulse fingerprinting is not complex enough to perfectly prevent spoofing attacks, likely because it is currently designed only for anti-interference purposes instead of security. However, it is not trivial to design a complex fingerprinting while ensuring eye safety.
\end{observation}

\nsubsubsection{Case Study on Specific LiDARs}
\label{sec:specific-lidars}
Using our measurement setup, we also found 5 uniquely-interesting characteristics of specific LiDARs: the use of rare wavelength in OS1-32, relay attack on Leddar Pixell, zero-distance sensing of XT32, extra-wide vertical FOV of Helios 5515, and simultaneous laser firing on OS1-32 and VLS-128, which led to 2 new findings that are different from existing understandings in the community. Details are in Appendix~\ref{appendix:specific-lidars}.

\nsubsection{Object Detector-Level Measurements (RQ3)} 
\label{sec:od_injection}

\nsubsubsection{Modeling of the Spoofing Attack Capability in Point Injection} \label{sec:od_injection_modeling}
Similar to prior works in this problem space~\cite{cao2019adversarial, jiachen2020towards, hallyburton2022security}, to enable large-scale measurements of the attack capabilities on the object detector side, we need to mathematically model the point injection capabilities. Prior efforts in such modeling did not systematically consider (1) the modeling of the CPI attack capability (\S\ref{sec:cpi}), and (2) common new-gen LiDAR features that can fundamentally affect the object injection attack capabilities such as timing randomization and pulse fingerprinting. Leveraging this measurement study, we can thus develop a new modeling that addresses both fronts: 

\vspace{-0.2in}
\footnotesize
\begin{align}
    \mathcal{P}_I (x_{ij}) = 
     x_{ij} + (\delta_{ij}^{\rm rand} + \delta^{\rm inner}_{ij} + \delta^{\rm inter}) \cdot g(x_{ij}), \ x_{ij} \in \mathcal{C}_n \subset \mathcal{C} 
     \label{eq:attack_cap}
\end{align}
\normalsize
, where $\mathcal{C}$ is a point cloud (i.e., chosen pattern) that the attacker originally wants to inject (e.g., the point cloud of a vehicle); 
$\mathcal{C}_n$ is a point cloud randomly downsampled to $n$ points from $\mathcal{C}$ to model the impact from pulse fingerprinting (\S\ref{sec:sec_enchance_feats});
$x_{ij} \in \mathbb{R}^{3}$ is a point injected by the attack at $i$-th altitude and $j$-th azimuth; $g: \mathbb{R}^{3} \rightarrow \mathbb{R}^{3}$ is a function to obtain a movable unit direction of point $x$. Due to the physics of LiDAR, each point can only move along with the laser direction. As the LiDAR typically locates at the origin of the point cloud, $g(x_{ij})$ can be written as $\frac{x}{||x||_{2}}$ in this case.
To model the spoofing inaccuracy and the effect of the timing randomization, we add 3 types of errors: $\delta_{ij}^{\rm rand}$, $\delta^{\rm inner}_{ij}$, and $\delta^{\rm inter}$. Randomization error $\delta_{ij}^{\rm rand}$ is the error introduced by the timing randomization. It follows a certain distribution based on the target LiDAR as measured in Table~\ref{tbl:randomization}. 
We use $\delta_{ij}^{\rm rand} \sim \mathcal{N}(0, \sigma$) for the Gaussian distribution and $\delta_{ij}^{\rm rand} \sim \mathcal{U}(-\frac{\max - \min}{2}, \frac{\max - \min}{2})$ for the uniform distribution. $\delta_{ij}^{\rm inner}$ and $\delta^{\rm inter}$ are the inner-frame and inter-frame errors, respectively. Based on the measurements in~\S\ref{sec:cpi_capability}, we sample $\delta_{ij}^{\rm inner}$ from $\mathcal{N}(0, 10 \ {\rm cm})$ and sample $\delta^{\rm inter}$ from $\mathcal{N}(0, 35 \ {\rm cm})$. Note that $\delta^{\rm inter} \in \mathbb{R}$ does not depend on each point $x_{ij}$, i.e., one scalar value is sampled per case and applied to all points in the case. To the best of our knowledge, this is the first modeling
of the point injection capability that can cover both first- and new-gen LiDAR features.

\newpart{Note that this modeling does not cover the impacts from the simultaneous laser firing feature (\S\ref{sec:sim_fireing}). This is because although such a feature can make the CPI attack capability infeasible, (1) it cannot prevent the point injection itself; (2) details of the simultaneous firing pattern are not well documented; and (3) the more recent LiDARs (e.g., the ones after 2019 in Table~\ref{tbl:target_lidars}) do not have such a feature. We thus leave its modeling to future work.}

\nsubsubsection{Experimental Setups} \label{sec:model_level_scenario}
We perform the experiments based on a scenario in the KITTI dataset.
Fig.~\ref{fig:synth_scenario} in Appendix %
depicts the generated scenario: we place a 3D vehicle object in front of the victim and move the corresponding points to the object's surface, following the same methodology used in~\cite{jiachen2020towards, cao2021invisible}. 
We generate 15 scenarios varying the distance between the victim to the vehicle object from 0 m to 14 m (0 m means the victim's nose touches the vehicle object's tail). 
Fig.~\ref{fig:n_points} in Appendix shows the number of points on the injected vehicle object.
In addition to the original point clouds, we evaluate point clouds downsampled from the original one to 10, 50, 100, and 200 points (i.e., $n = 10, 50, 100,$ and $200$ in Eq.~\ref{eq:attack_cap}) as a modeling of the fingerprinting effect  (\S\ref{sec:sec_enchance_feats}).
We judge the object injection as successful if there exists a detected object overlapping with the ground truth area of the injected object, i.e., the IoU between the ground truth and detected object is more than zero.

\vspace{0.05in}
\nsubsubsection{Results} \label{sec:od_injection_eval}  \label{sec:impact_model} \label{sec:impact_data} \label{sec:impact_noise}
~

\begin{figure}[t!]
\centering
\includegraphics[width=\linewidth]{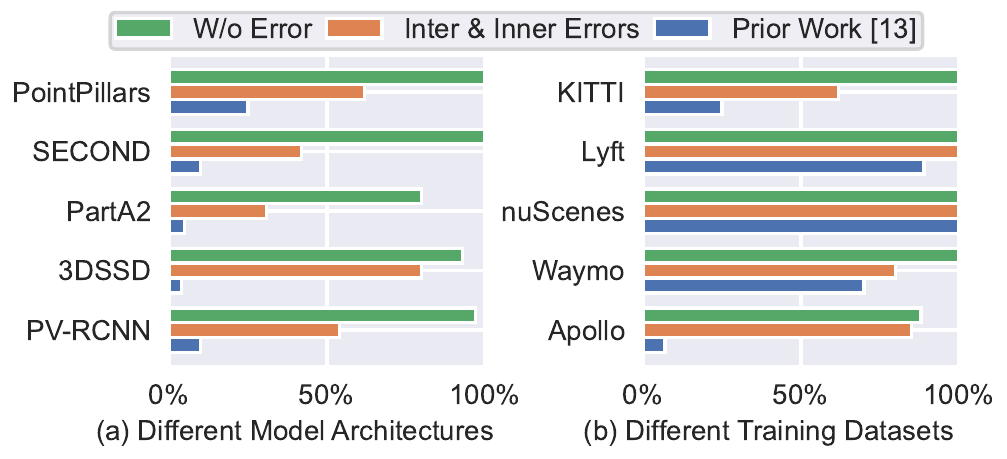}
\vspace{-0.2in}
\caption{
Object injection attack success rates under 3 different types of noises on (a) multiple models trained on the KITTI dataset and (b) PointPillars trained on different datasets.
}
\label{fig:obj_impact_errors}
\end{figure}

\textbf{Impact of Error Modeling on Spoofing Accuracy:}
We measure the impacts of 3 types of different modeling: without errors, our modeling with inner- and inter-frame errors (\S\ref{sec:cpi_capability}), and the error modeling used in the Frustum attack~\cite{hallyburton2022security}, which is the latest modeling effort in prior works and simply adds Gaussian errors along with the vehicle's Cartesian coordinates (forward, left, and up) following ($\mathcal{N}(1, 0.1)$, $\mathcal{N}(0, 0.5)$, $\mathcal{N}(1, 0.2)$) in meters, respectively.
Fig.~\ref{fig:obj_impact_errors} shows the attack success rates against our targeted models with different architectures and different training datasets (\S\ref{sec:target_lidar_od}).
As shown, the different error modeling can generally cause significant differences in the attack success rates, e.g., the results without errors and with the naive modeling in~\cite{hallyburton2022security} can differ 96\% and 70\% respectively on average to those using our modeling. Specifically, the latest modeling effort in~\cite{hallyburton2022security} significantly over-estimates the errors along their forward and up coordinates, which causes the model-level success rates to be generally lower than those with our more systematic modeling based on actual experiments (\S\ref{sec:cpi_capability}).

\begin{observation}{RQ1} 
Error modeling on the spoofing accuracy can significantly affect the object detector-level attack results. The latest prior efforts~\cite{hallyburton2022security} largely overestimate the errors as compared to our systematically modeled and quantified ones via experiments. 
\end{observation}
\vspace{-0.1in}

\begin{figure}[t!]
\centering
\includegraphics[width=\linewidth]{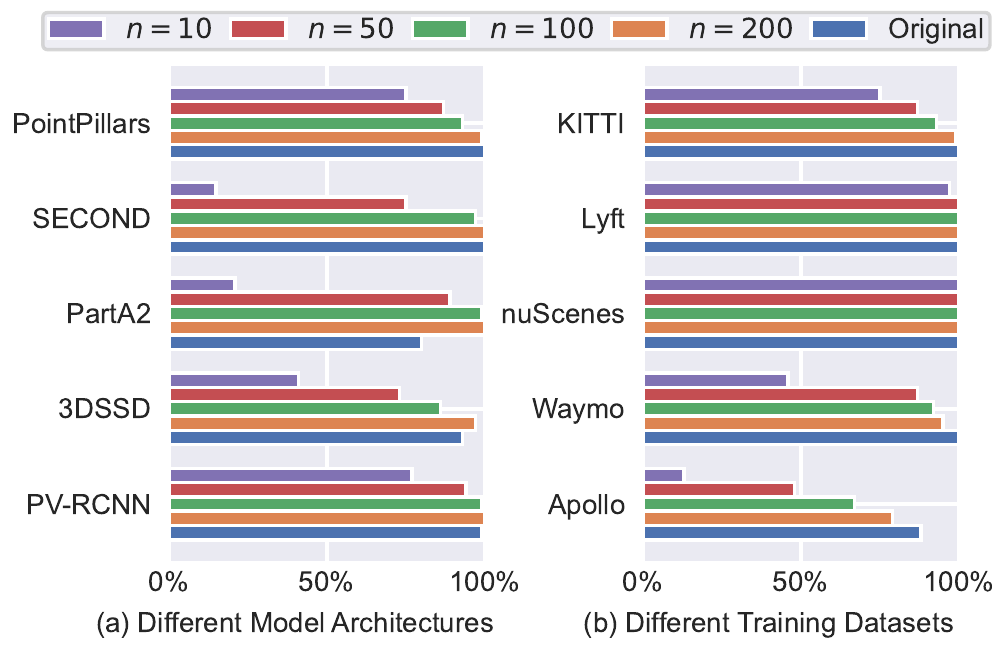}
\vspace{-0.2in}
\caption{
Object injection attack success rates under different down-sampling levels $n$ on (a) multiple models trained on the KITTI dataset and (b) PointPillars trained on different datasets.
}
\label{fig:obj_fingerprinting}
\end{figure}

\textbf{Impact of Pulse Fingerprinting:}
Fig.~\ref{fig:obj_fingerprinting} shows the attack success rates under different down-sampling levels $n$ as a modelling of different pulse fingerprinting levels (\S\ref{sec:sec_enchance_feats}). 
To perform control experiments of the fingerprinting impact factor, we do not apply inter- and inner-frame errors in this experiment. 
Generally, fingerprinting with higher complexity (i.e., lower $n$) shows higher defense capability. However, when $n=100$, which can represent the fingerprinting complexity level today (XT32 as measured in~\S\ref{sec:sec_enchance_feats}), the defense effectiveness is actually still very minimal (reduce the attack success rate only by 3\% on average). The defense effectiveness only becomes significant when $n$ reaches as low as 10, while such higher effectiveness is still model architecture and training dataset dependent.

\begin{observation}{RQ3}
If with enough complexity, pulse fingerprinting can indeed show high defense capability to object injection attacks (although can be object detector model architecture and training dataset dependent). Unfortunately, the complexity of pulse fingerprinting today may not be enough to achieve such high defense capability. 
\label{finding:pulsefp}
\end{observation}
\vspace{-0.1in}

\textbf{Impact of Timing Randomization:}
Table~\ref{tbl:inj_rand_arch} and~\ref{tbl:inj_rand_data} show the attack success rates under different timing randomization levels based on our measurements in~\S\ref{sec:sec_enchance_feats}. As shown, timing randomization can in general significantly reduce the attack success rates, e.g., the average attack success rates with randomization are dramatically reduced to at most 34\% for most models, while those without randomization are at least 88\%. We also measured the attack success rates with both timing randomization and pulse fingerprinting, with the fingerprinting level set at the observed level today ($n=100$, based on XT32 measurements in~\S\ref{sec:sec_enchance_feats}). However, consistent with Finding~\ref{finding:pulsefp}, we are not able to observe significant changes in the attack success rate reduction.

\begin{observation}{RQ3}
Timing randomization, even with those with low randomization entropy that is realized for anti-interference purposes instead of security today, can have significant defense capabilities against object injection attacks. Today's fingerprinting complexity level again may not be able to significantly help boost the defense capabilities when used together with timing randomization.
\end{observation}

Meanwhile, we further notice that the defense effectiveness with timing randomization can differ significantly when applied to models trained on different datasets. For example, as shown in Table~\ref{tbl:inj_rand_data}, the attack success rates are reduced by only 10\% and even 0\% on average across all randomization levels for the models trained on Lyft and nuScenes respectively, while the success rates can be reduced by as high as 84\% and 88.6\% for the same model design trained on Waymo dataset and the private dataset from Apollo. In particular, the models trained on Lyft and nuScenes are particularly vulnerable to object injection attacks since as shown in Fig.~\ref{fig:obj_fingerprinting}, these models can detect the vehicle even when the vehicle point cloud is randomly downsampled to as few as only 10 points. This suggests that the training dataset choice can play a significant role in determining the model vulnerability to spoofing attacks. Later in~\S\ref{sec:od_removal}, we will continue investigating this aspect together with the object removal attack measurement results.

\begin{table}[t!]
\centering
\footnotesize
\setlength{\tabcolsep}{1.3pt}
\setlength{\aboverulesep}{0pt}
\setlength{\belowrulesep}{0pt}
\renewcommand{\arraystretch}{0.75}
\caption{Object injection attack success rates under different randomization levels based on our measurements in~\S\ref{sec:sec_enchance_feats} on \textit{different model architectures}. $\emptyset$: No randomization. Avg.: The average results across all these randomization levels.
}
\label{tbl:inj_rand_arch}
\begin{tabular}{ccccccc}
\toprule
LiDAR   & Rand. model {[}m{]}         & PointPillars & SECOND & PartA$^{2}$ & 3DSSD & PV-RCNN \\\hline\hline
VLP-16  & $\emptyset$                 & \underline{100\%}        & \underline{100\%}  & 80\%   & 93\%  & 97\%    \\\hline\hline 
Helios  & $\mathcal{N}(0, 1.5)$       & 2\%          & 54\%   & 41\%   & 7\%   & 24\%    \\
L515    & $\mathcal{N}(0, 7.5)$       & \textbf{0\%}          & 24\%   & 14\%   & 7\%   & \textbf{0\%}     \\
Horizon & $\mathcal{U}(-45, 45)$      & 39\%         & 35\%   & 21\%   & 30\%  & 17\%    \\
OS1-32  & $\mathcal{U}(-58, 58)$ & 47\%         & 38\%   & 21\%   & 28\%  & 23\%    \\
Pixell  & $\mathcal{U}(-191, 191)$   & 60\%         & 21\%   & 20\%   & 8\%   & 43\%    \\ \hline
        & Avg.                         &   30\%         & 34\%   & 23\%   & 16\%  & 21\%  \\\hline\hline
  \multicolumn{3}{l}{With fingerprinting effect $n=100$:}        &     &     &    &     \\
                    & Avg.     & 38\%         & 20\%   & 16\%   & 18\%  & 41\%   \\\toprule
\end{tabular}
\end{table}

\begin{table}[t!]
\centering
\footnotesize
\setlength{\tabcolsep}{3.5pt}
\setlength{\aboverulesep}{0pt}
\setlength{\belowrulesep}{0pt}
\renewcommand{\arraystretch}{0.75}
\caption{Object injection attack success rates under different randomization levels based on our measurements in~\S\ref{sec:sec_enchance_feats} on \textit{models trained on different datasets}. $\emptyset$: No randomization.  Avg.: The average results across all these randomization levels.
}
\label{tbl:inj_rand_data}
\begin{tabular}{ccccccc}
\toprule
LiDAR   & Rand. model {[}m{]}         & KITTI & Lyft  & nuScenes & Waymo & Apollo \\ \hline\hline
VLP-16  & $\emptyset$                 & \underline{100\%} & \underline{100\%} & \underline{100\%}    & \underline{100\%} & 88\%   \\\hline\hline 
Helios  & $\mathcal{N}(0, 1.5)$       & 2\%   & \underline{100\%} & \underline{100\%}    & 26\%  & 48\%   \\
L515    & $\mathcal{N}(0, 7.5)$       & \textbf{0\%}   & 60\%  & \underline{100\%}    & \textbf{0\%}   & \textbf{0\%}    \\
Horizon & $\mathcal{U}(-45, 45)$      & 39\%  & 96\%  & \underline{100\%}    & 7\%   & \textbf{0\%}    \\
OS1-32  & $\mathcal{U}(- 58, 58)$ & 47\%  & 99\%  & \underline{100\%}    & 12\%  & \textbf{0\%}    \\
Pixell  & $\mathcal{U}(- 191, 191)$   & 60\%  & 96\%  & \underline{100\%}    & 36\%  & 1\%    \\ \hline
        & Avg.                         & 30\%  & 90\%  & \underline{100\%}    & 16\%  & 10\%   \\ \hline\hline
  \multicolumn{3}{l}{With fingerprinting effect $n=100$:}        &     &     &    &     \\
        & Avg.                         & 38\%  & 85\%  & 92\%    & 33\%  & 5\%   \\ \toprule
\end{tabular}
\end{table}

\nsection{Object Removal Attack Measurements} \label{sec:removal_attack}

After the measurements on object injection attacks, in this section we measure the other important class of LiDAR spoofing attacks: object removal attacks (\S\ref{sec:attack_taxonomy}). 

\nsubsection{LiDAR-Level Measurements (RQ1, RQ2)}\label{sec:lidar_removal}

Table~\ref{tbl:hfa} shows the LiDAR-level attack capability measurement results for the PRA attack and the newly-identified HFR attack (\S\ref{sec:high_freq_attack}) on different LiDARs. To count the removed points, starting from the attacked point cloud, we first identify the attack-induced spoofed points using the same method as in \S\ref{sec:lidar_injection} and remove them. Next, we subtract the remaining points in the attacked point cloud from the benign point cloud; now the remaining points in the benign point cloud are thus the removed benign points by the attack (detailed in Appendix~\ref{appndix:count_method}).
We select 1 MHz as the pulse frequency and 80V as the laser drive voltage for HFR (detailed in Appendix~\ref{appndix:hfr_freq}).

As shown in Table~\ref{tbl:hfa}, with our spoofer improvements in~\S\ref{sec:spoofer_design}, PRA can now achieve over 6,600 removed points on VLP-16, which is 65.5\% more than the original paper~\cite{cao2023you}. However, such strong attack capability is limited to the first-gen LiDARs; for the new-gen ones, PRA is not applicable to any of them since its requirement of synchronization is directly foiled by common next-gen LiDAR features such as timing randomization and pulse fingerprinting (\S\ref{sec:sec_enchance_feats}). 

\begin{observation}{RQ1}
Due to the requirement of synchronization, the basic design assumption of the latest object removal attack is generally broken for next-gen LiDARs due to common features such as timing randomization and pulse fingerprinting. 
\end{observation}

\begin{figure}[t!]
\centering
\vspace{0.1in}
\includegraphics[width=\linewidth]{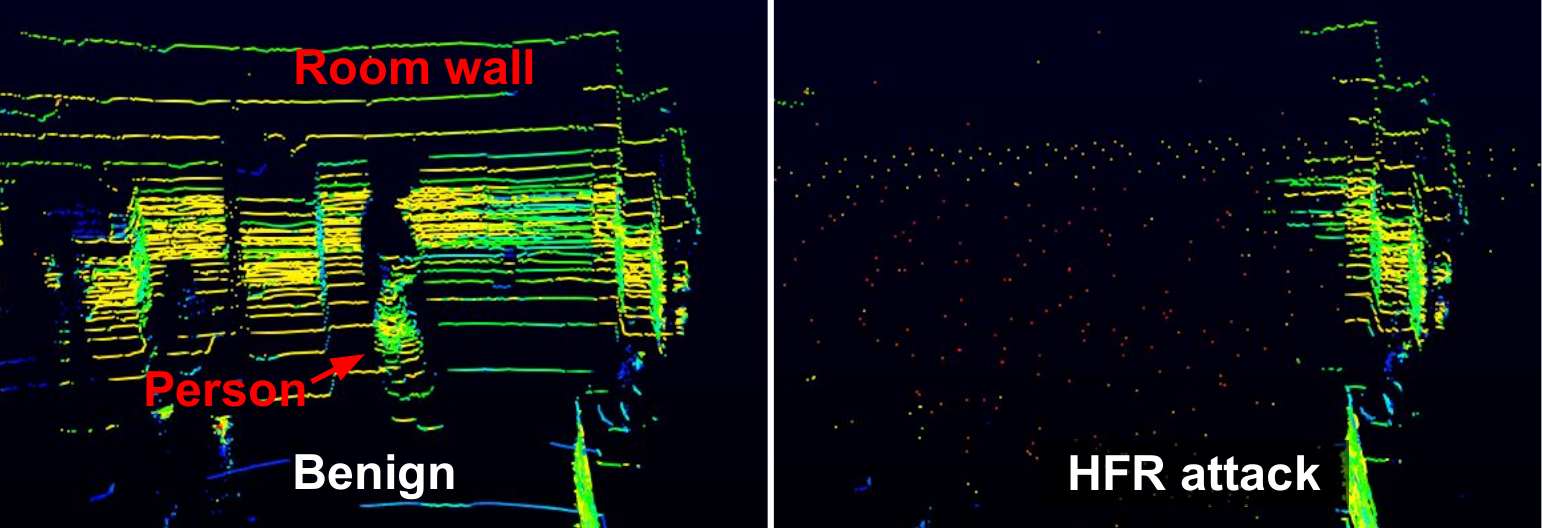}
\caption{HFR attack effect on VLP-32c. Patterns of a person and the majority of the room wall are completely removed. 
}
\label{fig:removal_attack}
\end{figure}

Since there is no need of synchronization, as shown in Table~\ref{tbl:hfa}, the newly-identified HFR attack can be generally applied to all LiDARs in the table, no matter if the targeted LiDAR is first- or new-gen. Specifically, for the first-gen ones, HFR is able to achieve comparable point removal capability to PRA, e.g., removing $>$5,300 points in $>$85$^{\circ}$. Fig.~\ref{fig:removal_attack} shows an example of the HFR attack effect on VLP-32c. As shown, the point cloud patterns of a person and the majority of the room wall are completely removed, with only some points with random noise patterns left.
Compared to PRA, the point removal success rate of HFR is slightly slower (72-78\% versus 80-97\% in PRA), which is expected since PRA has more precise control of points with synchronization. 

\begin{figure}[t!]
\centering
\vspace{-0.1in}
\includegraphics[width=\linewidth]{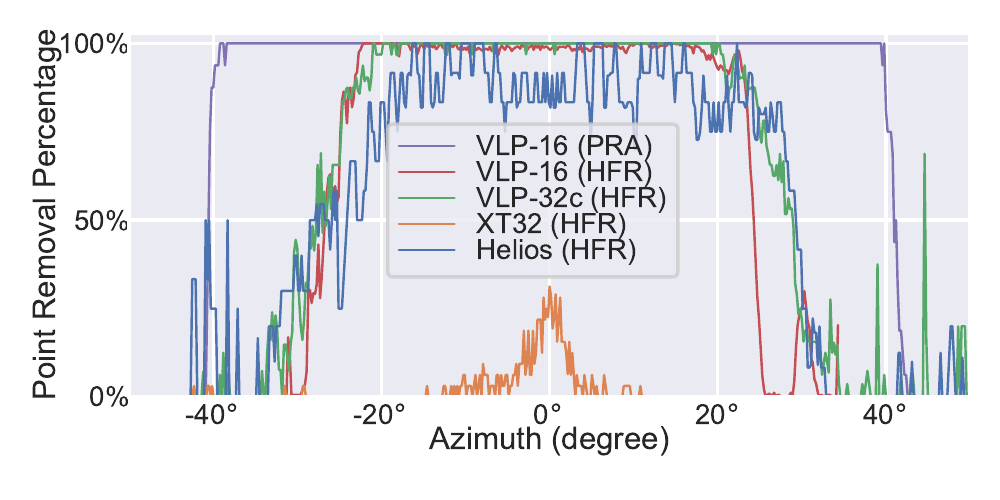}
\vspace{-0.3in}
\caption{Point removal percentage of PRA~\cite{cao2023you} and the HFR attack (\S\ref{sec:high_freq_attack}) for the azimuth angles under attack.%
}
\label{fig:hfa_asr}
\end{figure}

For next-gen LiDARs with timing randomization, HFR is still highly-effective since the high-frequency lasers can still hit all legitimate measurements within the affected azimuthal range despite the randomized timing of legitimate laser firing, showing the capability of removing 4k-206k points (for OS1-32 it is only 28 due to the lack of laser diodes with matching wavelength, explained in~\S\ref{sec:sec_enchance_feats}). For Helios, the point removal success rate is lower mainly due to its extra-wide vertical FOV (Appendix~\ref{sec:case_helios}); in the vertical range critical to real-world attack scenarios (e.g., 30-40$^{\circ}$ in the center), the success rate is similar to others (Fig.~\ref{fig:hfa_asr}). For next-gen LiDAR with pulse fingerprinting (XT32), the point removal capability is significantly reduced due to the difficulty for the attack pulses to coincide with the fingerprinting interval. However, as we found in~\S\ref{sec:sec_enchance_feats}, there are still chances to randomly remove $\sim$100 points by using a pulse frequency that is most likely to coincide with the fingerprinting interval.

\begin{observation}{RQ2}
\label{finding:hfr}
For new-gen LiDARs, although they are no longer generally vulnerable to the latest practical attacks such as PRA, they unfortunately still remain generally vulnerable to object removal attacks, with a similar level of practical attack capabilities as PRA, due to the possibility of new asynchronized removal attack designs such as HFR.
\end{observation}

\begin{table}[t!]
\centering
\footnotesize
\setlength{\tabcolsep}{1.2pt}
\setlength{\aboverulesep}{0pt}
\setlength{\belowrulesep}{0pt}
\renewcommand{\arraystretch}{0.9}
\caption{
LiDAR-level attack capability measurements for object removal attacks. PRA~\cite{cao2023you} is only applicable to the first-gen LiDARs. Symbols are the same as in Table~\ref{tbl:distance}.
}
\label{tbl:hfa}
\begin{tabular}{cc|cc|cccc|c}
\toprule
 &       & \multicolumn{2}{c|}{\multirow{2}{*}{First-Gen}} & \multicolumn{5}{c}{New-Gen}  \\  \cline{5-9}
 &       &  & & \multicolumn{4}{c|}{w/ Timing Randomization}    & w/ Fingerprint  \\  \cline{3-9}
 &       & VLP-16  & VLP-32c & OS1-32 & Helios    & Horizon  & L515    & XT32  \\  
 \hline  \hline
\multirow{3}{*}{\begin{tabular}[c]{@{}c@{}}PRA\\ \cite{cao2023you} \\ \end{tabular}} & $\mathcal{N}$     & 6,621    & 9,711    & N/A      & N/A       & N/A       & N/A & N/A     \\
                     & $\mathcal{R}$     & 96.9\% & 82.9\% & N/A      & N/A       & N/A       & N/A  & N/A    \\
                     & $\theta$ & 85.4$^{\circ}$ & 73.2$^{\circ}$ & N/A      & N/A       & N/A       & N/A & N/A     \\ \hline
\multirow{3}{*}{\begin{tabular}[c]{@{}c@{}}HFR\\(\S\ref{sec:high_freq_attack})\end{tabular}} & $\mathcal{N}$     & 5,358    & 8,778    & 28 & 4,108    & 19.2k    & 206k   & 113   \\
                     & $\mathcal{R}$     & 78.1\%  & 72.2\% & 43.8\%  & 24.8\%  & 79.9\%   & 91.3\%   & 2.1\% \\
                     & $\theta$ & 85.8$^{\circ}$ & 76.0$^{\circ}$ & 0.72$^{\circ}$  & 103.4$^{\circ}$  & 81.7$^{\circ}$ & 70.0$^{\circ}$ & 34.2$^{\circ}$ \\ \toprule
\end{tabular}
\raggedright

* N/A: Attack is not applicable to the LiDAR\\
\end{table}

\nsubsection{Object Detector-Level Measurements (RQ3)} \vspace{0.1in}

\nsubsubsection{Modeling of the Spoofing Attack Capability in Point Removal} \label{sec:od_removal_modeling}
Similar to the object injection attack side, we first perform a mathematical modeling of the spoofing attack capabilities in point removal in order to enable large-scale object detection-level attack capability measurements. For removal attacks, the point removal goal is the same for all point measurements that can be hit by the attack laser (e.g., move to within MOT for PRA, or place the point to a random position for HFR). Thus, the main factor that can affect the removal capability is whether the point is at an azimuth angle that can be effectively hit by the attack laser. For example, Fig.~\ref{fig:hfa_asr} shows the point removal percentage for the azimuth angles under attack. As shown, for VLP-16 under both PRA and HFR, the points in the center of the attack laser-hit azimuth range are almost 100\% removed regardless of altitudes and distances, while the removal percentages symmetrically decrease from the center to the side.

Based on these observations, we model the attack capability for point removal $\mathcal{P}_R$ as follows:
\vspace{-0.06in}
\begin{align}\footnotesize
    \mathcal{P}_R (x_{ij}) :=
    \begin{cases}
    \xi \cdot g(x_{ij}) & {\rm if } \  \operatorname{Bernoulli}(p_j) = 1 \\
    x_{ij} & {\rm othewise } 
  \end{cases}
     \label{eq:attack_modeling_removal}
\end{align}\vspace{-0.0in}\normalsize, where $x_{ij} \in \mathbb{R}^{3}$ is its point at $i$-th altitude and $j$-th azimuth. For each point, we decide whether it will be removed or not based on a Bernoulli trial with a per-azimuth probability $p_j$, which can be set using the measured point removal percentage as in Fig.~\ref{fig:hfa_asr}. $\xi$ models the removal effects for both HFR and PRA. For HFR, 
$\xi$ is the error of the HFR attack, which distributes the original points to random locations along with the laser direction $g(x_{ij})$. 
Thus, $\xi$ can be written as 
$\xi := \mathcal{U} \left ( 0, \frac{c}{2} \cdot \frac{1}{f} \right )$, where $\mathcal{U}(a, b)$ is the uniform distribution with the minimum $a$ and maximum $b$ values, $c$ is the speed of light, $f$ is the frequency of HFR attack pulses. This is derived from the fact that the maximum ToF is capped by the time interval of the high-frequency attack pulses in HFR. For PRA, $\xi$ is set to 0 since the point removal principle of PRA is to move the points toward the origin as much as possible to locate them within MOT (\S\ref{sec:attack_taxonomy}). 
After applying Eq.~\ref{eq:attack_modeling_removal}, if $\mathcal{P}_R (x_{ij})$ is within the MOT or outside of the maximum detection range (Table~\ref{tbl:target_lidars}), point $x_{ij}$ is removed from the point cloud. To the best of our knowledge, this is the first mathematical modeling of the point removal capability for LiDAR spoofing attacks.

\vspace{0.1in}
\nsubsubsection{Model-Level Attack Capability Measurement}
\label{sec:od_removal}
~

\textbf{Experimental Setup.}\label{sec:model_level_scenario_removal}
We use the same evaluation scenarios as in~\S\ref{sec:model_level_scenario} (i.e., 15 scenarios varying the distance between the victim to the vehicle object from 0m to 14m). We consider the object removal as successful if the IoU between any detected objects and the ground truth object is 0. When applying $\mathcal{P}_R(.)$, we set $p_j$ using the measured point removal percentages for different LiDARs as in Fig.~\ref{fig:hfa_asr}. Note that for Helios, since its vertical FOV is much wider than others (Appendix~\ref{sec:case_helios}), we only calculate the point removal percentage for the FOV range related to our attack scenario (i.e., 33$^{\circ}$, which is enough to cover the height of the front vehicle to remove).

\begin{figure}[t!]
\centering
\includegraphics[width=\linewidth]{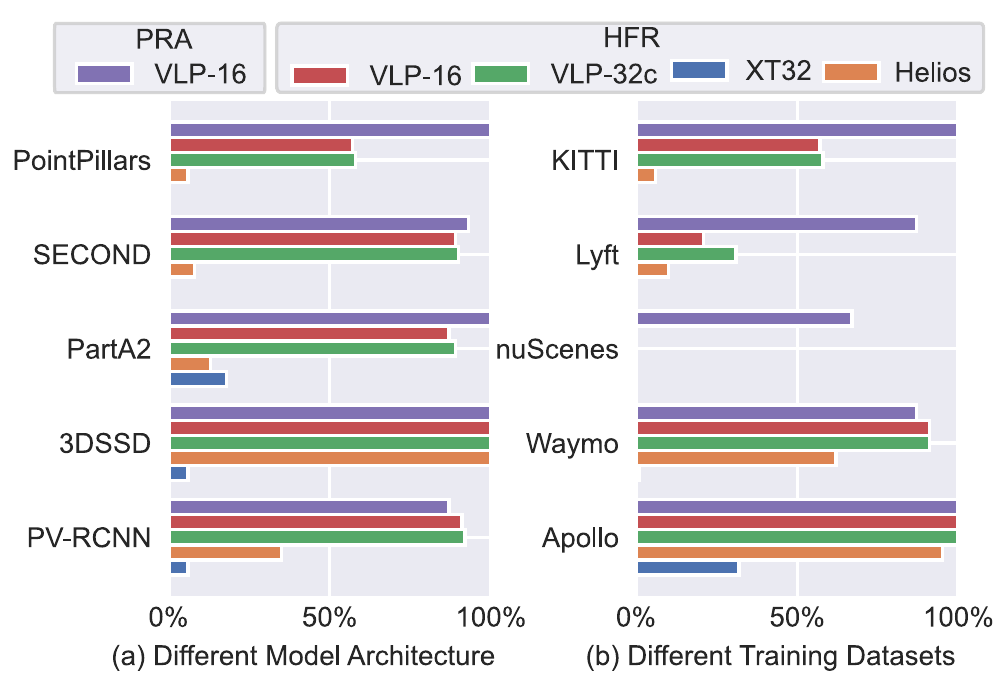}
\vspace{-0.2in}
\caption{
Object removal attack success rates of HFR and PRA for (a) 5 different models trained on the KITTI dataset and (b) PointPillars trained on 5 different datasets. %
}
\label{fig:removal_noise_data} \label{fig:removal_noise_arch}
\end{figure}

\begin{figure}[t!]
\centering
\includegraphics[width=\linewidth]{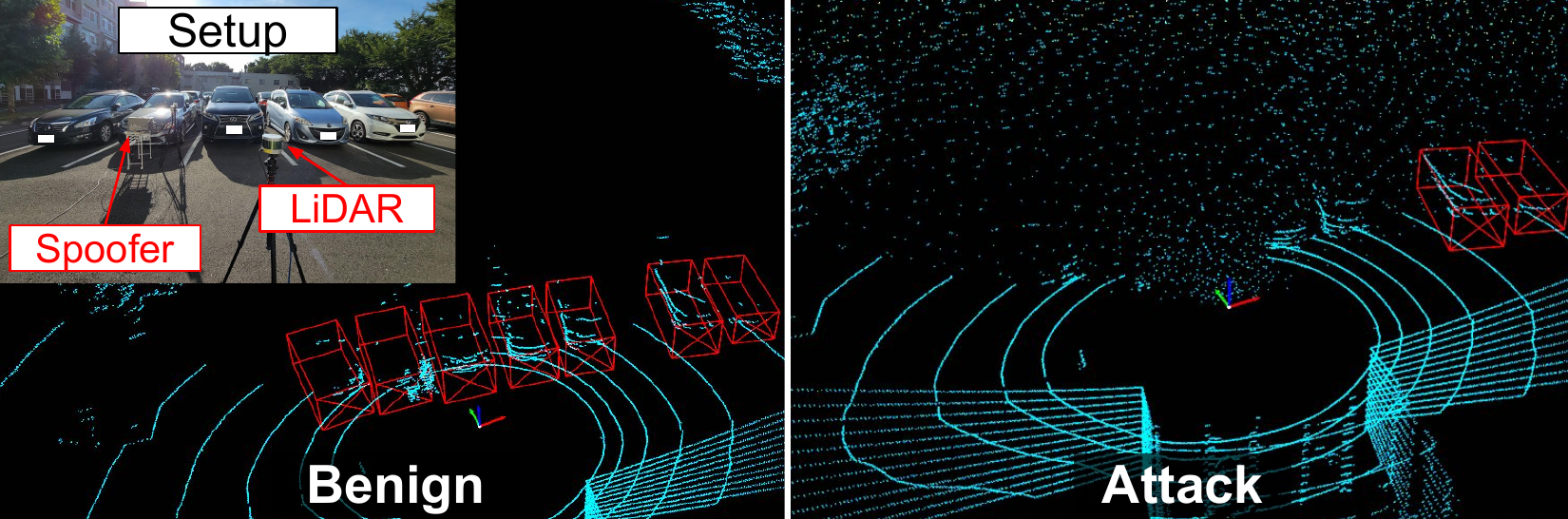}
\caption{Object removal attack effect against real vehicles using the HFR attack. The 5 front vehicles become undetected with a 100\% success rate for over 10 seconds (100 frames in total) by PointPillars~\cite{lang2019pointpillars} in Apollo~\cite{apollo}.
}
\label{fig:hfa_real_car}
\end{figure}

\label{sec:od_removal_eval}
\textbf{Results.} Fig.~\ref{fig:removal_noise_data} shows the object removal attack success rates for object detectors with different architectures and datasets. As shown, for the majority of the cases, HFR can reach similar success rates as PRA, which thus further extends our Finding~\ref{finding:hfr} to the object detector level. As a validation of such attack effectiveness in the physical world, Fig.~\ref{fig:hfa_real_car} shows the object removal attack effect against real vehicles using the HFR attack. As shown, the HFR attack is found to cause  5 front vehicles to become undetected by the Apollo model~\cite{apollo}, with 100\% success rate for over 10 seconds. In the figure, we can see the spatial features of the vehicles were completely destroyed (with some random points left) and thus no objects were detectable in the attacked region.

For new-gen LiDAR features, similar to the object injection side, timing randomization (Helios in Fig.~\ref{fig:removal_noise_data}) can significantly reduce the attack success rate in the majority of the cases (by 35\% on average from VLP-16). However, pulse fingerprinting shows much higher defense effectiveness compared to the object injection side: with the pulse fingerprinting strength level of XT32 (i.e., $\sim$100 randomly removed points), the average attack success rate is significantly reduced by 63\% on average from VLP-16, while such a reduction is 3\% on average on the object injection side (Fig.~\ref{fig:obj_fingerprinting}). 
\newpart{This is due to the trade-off of the object detector’s sensitivity to object injection and removal attacks. Since fingerprinting only allows a random subset of 100 points to be injected, the object detectors that are more vulnerable to object injection are those that are very sensitive to even a heavily-occluded point cloud pattern with a very small number of object points.}

\begin{observation}{RQ3}
Both timing randomization and pulse fingerprinting show high defense capabilities against object removal attacks in general; the defense capability from pulse fingerprinting is especially strong when compared with that against object injection attacks.
\end{observation}

In addition, similar to our findings on the object injection attack side, the model robustness to object removal attacks shows a high diversity across different model architectures and different training datasets, which is especially prominent across training datasets. Interestingly, the model robustness properties are generally the \textit{opposite} across object injection and removal attacks: for example, the models trained on Waymo and by Apollo are found to be much more robust against object injection attacks when compared to those trained on Lyft and nuScenes (\S\ref{sec:od_injection}), while the former ones become much less robust than the latter ones for object removal attacks. This might be because different dataset has different balances of the trade-off between false positives and more false negatives for the corner cases. As shown in Fig.~\ref{fig:obj_fingerprinting}, models trained on Lyft and nuScenes can have 100\% detection rate of a vehicle even when the point cloud is randomly downsampled to as few as 10 points. This can allow the trained models to have a very low false negative rate even for highly-occluded legitimate vehicle point cloud patterns, which can thus make the models highly robust to object removal attacks, but this also makes them highly vulnerable to object injection attacks.

\begin{observation}{RQ3}
The model robustness to object injection and removal attacks can be highly diverse when trained on different datasets. The choice of training datasets can incur large trade-offs between the model robustness to object injection attacks and that to object removal attacks.
\end{observation}

\label{sec:case_study}

\nsubsubsection{System-Level Attack Capability Measurement}
\label{sec:od_removal_system}
Due to the downstream components such as object tracking in real-world applications such as autonomous driving (AD), object detector-level object removal effect may not directly imply system-level attack effect such as vehicle crashes~\cite{jia2020fooling, shen2022sok}. Thus, in this section we further measure the attack capabilities of different object removal attacks at the system level, with the focus on the representative application: AD.

\textbf{Experimental Setup.} \label{sec:system_level_scenario}
We use a common setup widely used in previous work~\cite{cao2019adversarial, jiachen2020towards, hallyburton2022security, cao2023you}. For the AD system, we use Baidu Apollo 7.0~\cite{apollo}. For the driving simulator, we use LGSVL~\cite{lgsvl}. Both systems are known as industry-grade software.
Fig.~\ref{fig:overview_sim} in Appendix illustrates the experiment scenario. 
We place a sedan vehicle as a target object 200 m away from the victim AD and make the victim AD drive toward the target. We add up to 1 m random perturbation both laterally and longitudinally to (1) the starting point of the victim and (2) the target object. Before reaching the target object, the AD vehicle reaches and keeps 40 km/h. 
We set an attack start distance $\mathcal{D}$ and only start to apply the simulated removal attack effect when the distance from the victim AD vehicle to the front vehicle is $\leq\mathcal{D}$.
We use the collision rate over 10 trials as the evaluation metric, i.e., how many times the victim vehicle collides with the sedan out of 10 trials.

\begin{figure}[t!]
\centering
\includegraphics[width=\linewidth]{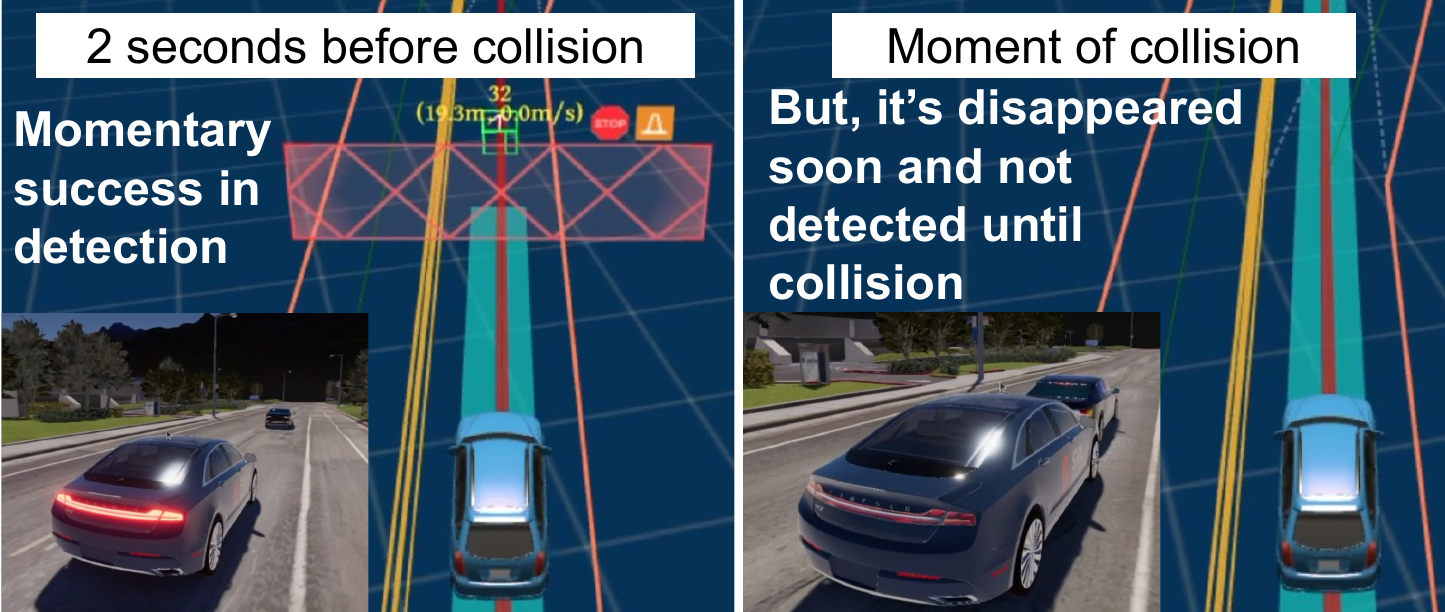}
\caption{An example HFR attack trial with $\mathcal{D}$ = 15 m on Helios. The remaining points are occasionally detected as an object but are not sufficient to avoid a collision.}
\label{fig:sim_attack}
\vspace{-0.05in}
\end{figure}

\textbf{Results.}
Table~\ref{tbl:sim_hfa} shows the collision rate over 10 trials for different LiDARs. As shown, although HFR has relatively weaker attack capabilities than PRA at raw point cloud (\S\ref{sec:lidar_removal}) and object detector (\S\ref{sec:model_level_scenario_removal}) levels due to the lack of synchronization, their attack capabilities at the system level are almost the same at every attack start distances $\mathcal{D}$, and reach 100\% collision rate when $\mathcal{D}$ is over 18 m. However, due to the requirement of synchronization, PRA is only applicable to first-gen LiDARs such as VLP-16, while HRA can be generally applied to new-gen ones such as Helios with similar attack effectiveness. Fig.~\ref{fig:sim_attack} shows an example attack trial under the HFR attack with $\mathcal{D}$ = 15 m on Helios. The target sedan can be momentarily detected before the collision, but the detection disappears soon and the sedan becomes undetected until and also after the collision.

For next-gen LiDAR with pulse fingerprinting (XT32), although HFR can still have $\leq$32\% model-level attack success rate (\S\ref{sec:od_removal_modeling}), it cannot achieve any system-level attack success at all (0\%), which is likely because the object removal rate is not high enough to fool object tracking~\cite{jia2020fooling}. This suggests that pulse fingerprinting can be quite effective in defending against object removal attacks at the system level. videos of these experiments can be found at our website~\cite{project_page}.

\begin{observation}{RQ3}
Although the HFR attack has weaker attack capabilities than PRA at point cloud and object detector levels, the system-level attack effect is almost the same. Since the HFR attack does not require synchronization, this means that real-world LiDAR applications such as autonomous driving, no matter if using first- or new-gen LiDARs, are facing practical threats from object removal attacks. Meanwhile, pulse fingerprinting can be quite effective in defending against the removal attack effect at the system level.
\end{observation}
\vspace{-0.13in}

\begin{table}[t!]
\centering
\footnotesize
\caption{Vehicle collision rates over 10 trials using PRA and HFR for different LiDARs with varying attack start distances. \textbf{Bold} and \underline{underline} highlight %
100\% and 0\% collision rates.}
\label{tbl:sim_hfa}
\setlength{\tabcolsep}{3.4pt}
\renewcommand{\arraystretch}{0.8}
\begin{tabular}{cccccccccc}
\toprule
      &       & Benign & 10m & 15m & 16m & 17m  & 18m  & 19m  & 20m  \\ \hline
PRA & VLP-16 & \underline{0/10}   & \underline{0/10} & 5/10 & 8/10 & 9/10 & \textbf{10/10} & \textbf{10/10} & \textbf{10/10} \\\hline
   & VLP-16 & \underline{0/10}   & \underline{0/10} & 6/10 & 7/10 & 8/10  & \textbf{10/10} & \textbf{10/10} & \textbf{10/10} \\
HFR & VLP-32c & \underline{0/10}   & 1/10 & 9/10 & 8/10 & \textbf{10/10} & \textbf{10/10} & \textbf{10/10} & \textbf{10/10} \\
   & XT32 & \underline{0/10}   & \underline{0/10} & \underline{0/10} & \underline{0/10} & \underline{0/10}  & \underline{0/10}  & \underline{0/10}  & \underline{0/10}  \\ 
   & Helios & 	\underline{0/10} & 	\underline{0/10} & 	6/10 & 	5/10 & 	\textbf{10/10} & 	\textbf{10/10} & 	\textbf{10/10} & 	\textbf{10/10}\\
\toprule
\end{tabular}
\end{table}

\nsection{Discussions} \label{sec:discussion}
\nsubsection{Defense Discussions}
\nsubsubsection{Sensor-Level Defenses}
\label{sec:sensor-level-defense}
Through large-scale measurements, we identify the defense capability of popular security-related features in new-gen LiDARs on the spoofing attack capabilities.
Table~\ref{tbl:defence} summarizes our observations regarding their defense effectiveness and limitations \newpart{against object injection and removal attacks}.

\begin{table}[t!]
\centering
\footnotesize
\caption{Defense effectiveness and limitations of security-related features in the new-gen LiDARs \newpart{against object injection and removal attacks}. \textbf{Bold} means desired properties; \underline{Underline} means undesire ones.
}
\label{tbl:defence}
\setlength{\tabcolsep}{3pt}
\setlength{\aboverulesep}{0pt}
\setlength{\belowrulesep}{0pt}
\renewcommand{\arraystretch}{0.9}
\begin{tabular}{ccc|ccc}
\toprule 
                  & \multicolumn{2}{c|}{Effectiveness} & \multicolumn{3}{c}{Limitations}        \\ \cline{2-6} 
Features           & Injection             & Removal           & Eye safety & Latency         & Range$\downarrow$* \\ \hline
Timing Random.    & \textbf{High}                & \textbf{High}                 & \textbf{No risk}    & \textbf{Low impact}      & \textbf{None}     \\
Pulse Fingerprint & Mid                & \textbf{High}                 & \underline{High risk} & Mid impact & \underline{High}  \\ 
Simul. Firing     & \underline{Low}                & \underline{None}                  & \textbf{Low risk}   & \textbf{Low impact}      & \textbf{Low}    \\\toprule
\end{tabular}
\raggedright

* Range$\downarrow$: Degradation of the effective sensing range of LiDAR
\end{table}

\textbf{Defenses against Object Injection Attacks.} 
As shown in \S\ref{sec:od_injection_eval}, the timing randomization feature has high mitigation capabilities, particularly on several models such as the industry-grade Apollo model. 
\newpart{Even with low randomization entropy, the timing randomization shows high defense capabilities. Meanwhile, higher randomization entropy has a higher defense capability. We suggest that the magnitude of the randomization be set as high as possible, e.g., std. dev. $\sigma\geq0.75$ %
}. 
Pulse fingerprinting also has mitigation capabilities but the complexity of fingerprinting in new-gen LiDARs today is not enough to effectively defend against injection attacks.
\newpart{Finally, simultaneous firing also has a defense effect as it can prevent the CPI attack capability (\S\ref{sec:sec_enchance_feats}). However, this alone is unlikely to affect the object injection attack goal in general since the attacker is still able to inject an object point cloud with at least the same number of points as the chosen pattern (but just cannot control the pattern very precisely).}

\textbf{Defenses against Object Removal Attacks.}
As discussed in~\S\ref{sec:od_removal_eval}, both timing randomization and pulse fingerprinting show high mitigation capability against removal attacks, especially pulse fingerprinting in XT32 shows high defense capability even against the HFR attack.
\newpart{The object detectors are typically tuned to avoid false negatives rather than false positives, especially for AD scenarios. This trade-off benefits pulse fingerprinting for defending against object removal attacks because the majority of object points are not compromised.
However, this tuning is a backfire to defend against object injection attacks because it allows a hundred injected points to be detected as an object.
}
However,  the current XT32-level pulse fingerprinting is still not enough to prevent all attacks as the HFR attack still has 32\% attack success rate on the Apollo model (\S\ref{sec:od_removal_eval}).

To improve this further, more complex fingerprint encoding is needed. However, this is fairly non-trivial due to the dilemma between eye safety and detection range.
More specifically, the most direct way to increase the fingerprint coding complexity is to increase the number of pulses for each distance measurement. However,  the laser power per unit time is capped to ensure eye safety. For example, if we shoot $N$ pulses for each point measurement, the power for each pulse should be $1/N$, and it will roughly degrade the effective sensing range by $1/\sqrt{N}$. To address the dilemma, future work can explore: (1) possible fingerprint coding designs both with high complexity and fewer pulses, and (2) use a wavelength to which the human eye is highly resistant, such as 1550 nm wavelength. \newpart{For the simultaneous firing feature, for object removal it does not have any defense capability since it just makes the attack effects on some sub-group of the points in the chosen pattern equal instead of having a removal effect of them.}

\newpart{\textbf{Summary.} Overall, the timing randomization and pulse fingerprinting features show promising defense effectiveness on at least one of the object detector-side attack goals. Particularly, timing randomization has high effectiveness on both attack goals while suffering from the least limitations in eye safety, latency, and sensing range, despite the simplicity of the method.
Thus, we strongly recommend implementing it in future LiDARs as a highly cost-effective measure to improve their resiliency against LiDAR spoofing attacks.}

\nsubsubsection{Software-Level Defenses}

Recently, several software-level defenses have been proposed against injection attacks. CARLO~\cite{jiachen2020towards}, SVF~\cite{jiachen2020towards}, and Shadow-Catcher\cite{hau2021shadow} leverage an assumption that the spoofer cannot directly spoof a point cloud pattern with sufficient number of points and precision to make it indistinguishable from a benign one.
However, we find that this assumption does not hold since our improved spoofer can achieve the CPI attack capability with a much larger number of points that is sufficient to directly spoof the complete point cloud pattern of a near-front vehicle (\S\ref{sec:inj_vlp16}).
Fortunately, VLP-16~\cite{VLP16} is the only LiDAR for which such CPI attack capability is feasible (\S\ref{sec:inj_other_lidars}). Thus, future research can focus more on more recent LiDARs, especially the new-gen ones, to explore software-level defense design possibilities. 
For object removal attacks, we still do not have effective defenses since PRA~\cite{cao2023you} is discovered very recently.
A potential software-level defense for the HFR attack is to detect a unique characteristic of the HFR attack that will cause a randomized point distribution pattern (like salt-and-pepper noise) in the area under attack, which is caused by the attack design (\S\ref{sec:high_freq_attack}) and can rarely occur in benign cases.

\newpart{
\nsubsection{Limitations of Our Study} \label{appendix:limitation}

\nsubsubsection{Aiming at Driving AD Vehicles}
In this study, we do not discuss the deployability of LiDAR spoofing attacks against real-world applications such as AD, especially when the victim AD vehicle is moving in high speed. We focus more on investigating the security properties of different types of LiDARs, rather than the system-level security analysis of autonomous driving in the real world. In a future study, we plan to demonstrate attacks against driving vehicles. %
A recent study~\cite{cao2023you} has demonstrated a system capable of victim LiDAR tracking and spoofer aiming to address this problem.
Therefore, we expect that targeting a driving vehicle is feasible. %

\nsubsubsection{LiDAR Model Coverage}
While we cover popular LiDARs as many as possible with our best efforts, it is infeasible to cover all public LiDARs due to our budget and their supply capacity. For example, we cannot cover 1550 nm LiDARs such as AEye~\cite{Aeye} and Luminar~\cite{Luminar} which utilize unique technology, e.g. adaptive scanning and FMCW ToF, and thus potentially have different spoofing capabilities.
We also cannot cover private LiDARs such as those in Waymo's robotaxi~\cite{waymoone} since they are not publicly available on the market.

\nsubsection{Considerations for Safe Experiments} \label{appendix:safety}
All experiments were conducted in a controlled environment and we wore safety goggles for extra eye safety. Note that the unit-area peak power of our laser is actually weaker than prior works~\cite{cao2023you} as we use a 50\% larger aperture lens. %

}

\nsection{Conclusion} \label{sec:conclusion}

In this work, we conduct the first large-scale measurement study on LiDAR spoofing attack capabilities on object detectors with 9 popular LiDARs, covering both first- and new-gen LiDARs, and 3 major types of object detectors trained on 5 different datasets. To facilitate the measurements, we (1) identify spoofer improvements that significantly improve the latest spoofing capability, (2) identify a new synchronized object removal attack (HFR)), and (3) perform novel mathematical modeling for both object injection and removal attacks. Through this study, we are able to uncover a total of 15 novel findings, including not only completely new ones due to the measurement angle novelty, but also many that can directly challenge the latest understandings in this problem space. We discuss defenses based on the findings. We hope that our findings can inspire and facilitate future security research on LiDAR spoofing, especially those targeting safety-critical application domains such as autonomous driving.

\vspace{-0.06in}
\section*{Acknowledgements}
\vspace{-0.06in}
This research was supported in part by the NSF CNS-1932464, CNS-1929771, CNS-2145493, USDOT UTC Grant 69A3552047138, JST SPRING JPMJSP2123, JST PRESTO JPMJPR22PA, and JSPS KAKENHI 21K20413.
\vspace{-0.06in}

{\small
\bibliographystyle{IEEEtran}
\bibliography{main.bib}
}

\begin{figure}[h!]
\centering
\includegraphics[width=\linewidth]{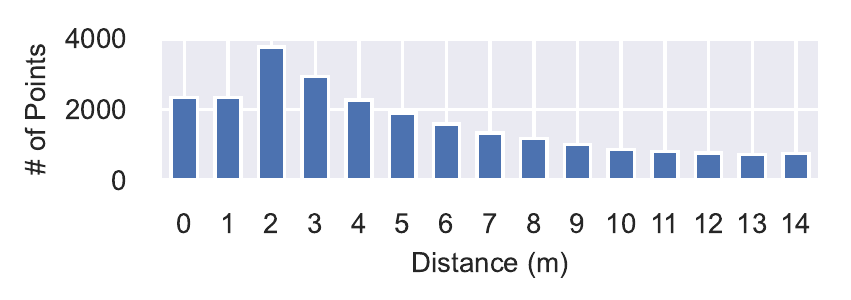}
\vspace{-0.25in}
\caption{The number of points for the targeted vehicle at each distance to the victim ego vehicle.}
\label{fig:n_points}
\vspace{-0.1in}
\end{figure}

\begin{figure}[h!]
\centering
\includegraphics[width=\linewidth]{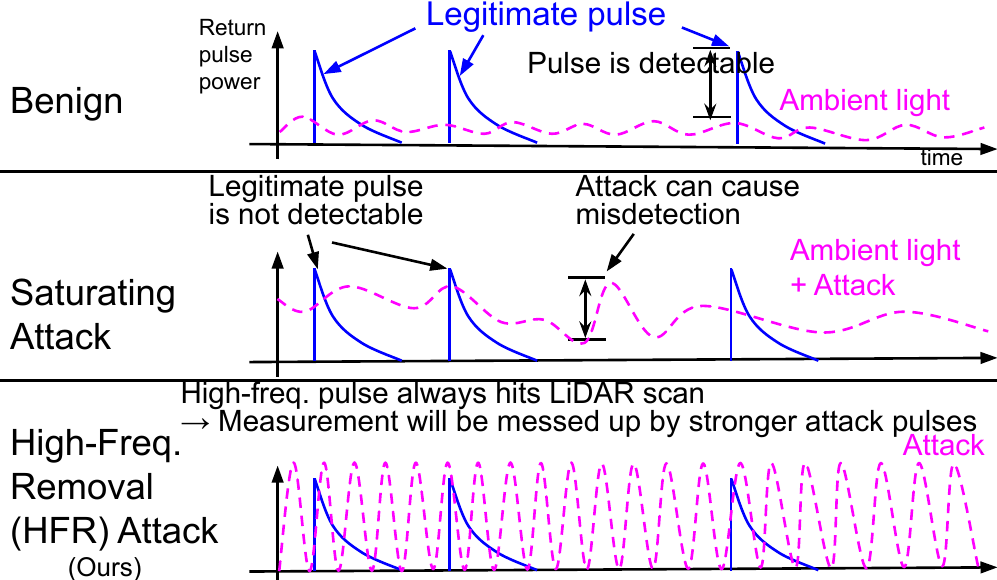}
\caption{Attack mechanism difference between the saturating attack and the newly-identified HFR attack (\S\ref{sec:high_freq_attack}).
}
\label{fig:highfreq_vs_saturating}
\end{figure}

\begin{figure}[h!]
\centering
\includegraphics[width=\linewidth]{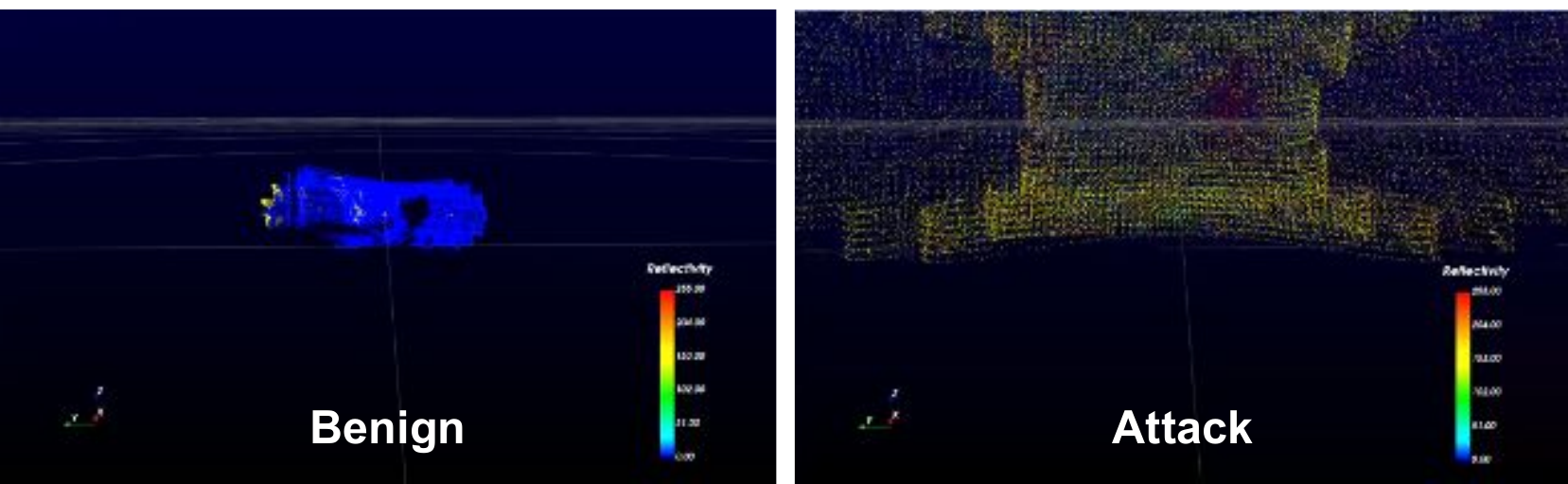}
\caption{Point cloud of Livox Horizon in benign and attack scenarios. The point cloud is totally randomized by the attack and the pattern of the rectangle area (our lab room) in the left figure completely disappears as shown in the right figure.}
\label{fig:livox_spoofing}
\end{figure}

\begin{figure}[h!]
\centering
\includegraphics[width=\linewidth]{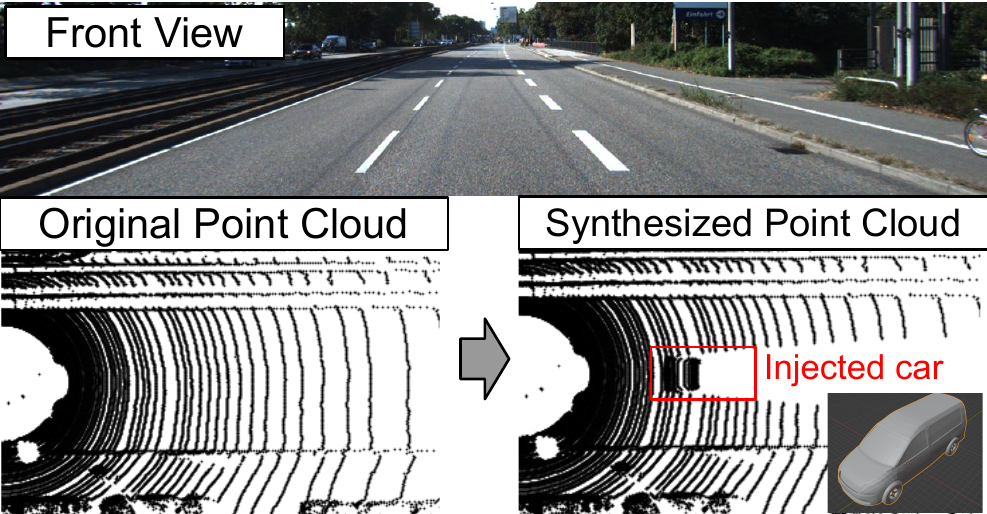}
\caption{Targeted experiment scenario from KITTI~\cite{Geiger2012CVPR}.}
\label{fig:synth_scenario}
\end{figure}

\begin{figure}[h!]
\centering
\includegraphics[width=\linewidth]{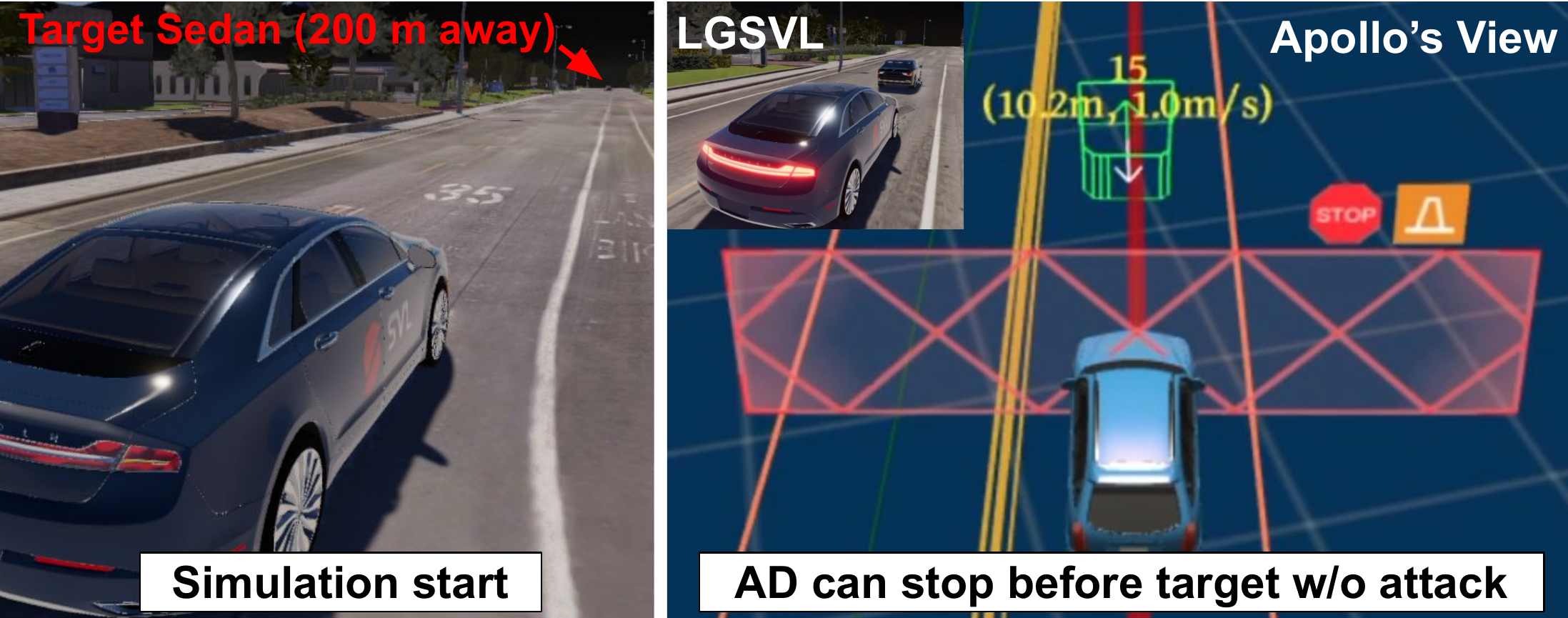}
\caption{Illustration of the evaluation scenario. AD starts driving from 200 m away and it can always successfully stop before the target sedan in the scenarios without attack.}
\label{fig:overview_sim}
\end{figure}

\newpage

\vspace{0.3in}
\appendices

\vspace{-0.2in}
\nsection{Detailed Explanations on Synchronized LiDAR Spoofing Attacks}
\label{appndix:sync_attack}

\begin{figure}[h!]
\centering
\vspace{0.05in}
\includegraphics[width=\linewidth]{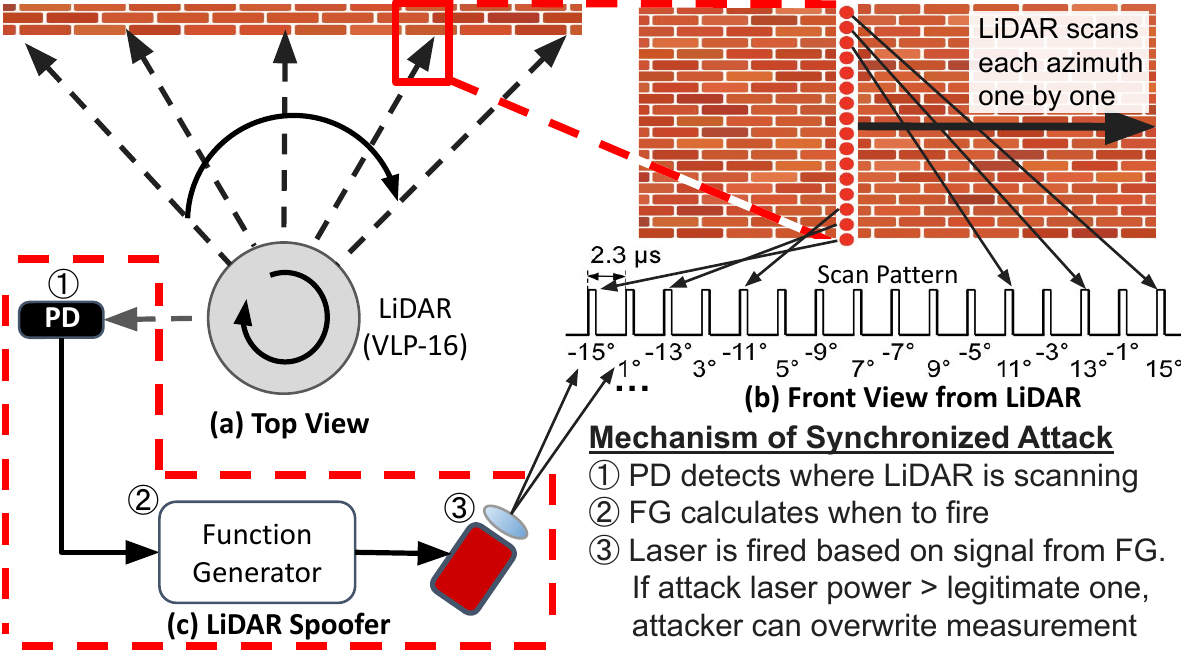}
\caption{Illustration of synchronized attacks on VLP-16. VLP-16 scans each azimuth (every 0.1$^{\circ}$) one by one. At an azimuth, it fires 16 lasers vertically based on the pre-defined scan pattern. Once attackers can identify its state by PD, they can know \textit{when} to fire a malicious laser based on the pattern.}
\label{fig:overview_sync}
\end{figure}

Fig.~\ref{fig:overview_sync} illustrates the synchronized attacks on VLP-16. The attack mechanism is common on both the injection attacks~\cite{shin2017illusion, cao2019adversarial, jiachen2020towards, hallyburton2022security} and the removal attacks~\cite{zhongyuan2021object, cao2023you} since their difference is whether they move points at target locations or move points into undetectable area. The attack procedure consists of 3 steps:~\circled{1} PD first receive the legitimate laser from the target LiDAR to know when the LiDAR will scan the point that the attacker want to change; ~\circled{2} FG plans when to fire lasers based on the information from PD and the pre-defined scan pattern of the target LiDAR;~\circled{3} the laser is fired based on the plan from FG through the gate driver and LD. Therefore, to achieve the CPI attack capabilities, the attacker must know exactly where LiDAR is scanning and the scan schedule must be predefined or predictable. The timing randomization breaks the assumption by randomizing the scan schedule.

\vspace{-0.05in}
\nsection{Details of LiDAR Spoofer Used in Experiments}
\label{appendix:spoofer}

Our spoofer operates in 4 steps (shown in Fig.~\ref{fig:spoofer1}): \circled{1} The photodetector (PD) receives a laser pulse from LiDAR; \circled{2} the transimpedance amplifier (TIA) amplifies the pulse to 100 mV to feed it into the function generator (FG); \circled{3} FG synchronizes with the LiDAR spanning pattern based on the received pulse and generates laser pulses according to the points that the attacker wants to inject; \circled{4} The gate driver (GD) drives the laser diode (LD) which transmits a laser pulse to the LiDAR.

We use the same PD (S6775~\cite{S6775}) and TIA (TL082~\cite{TL082}) as the prior work~\cite{cao2023you}. For other components, we used the equivalent or enhanced devices: FG is Agilent 81160A~\cite{81160A}, GD is EPC9126HC~\cite{EPC9126HC}, and LD is SPL PL90\_3~\cite{SPLPL90_3}. In particular, we utilize a more functional FG (Agilent 81160A) than the one used in previous studies (Tektronix AFG320~\cite{AFG320}) to better facilitate the exploration of the CPI attack capability.

\begin{figure}[h!]
\centering
\includegraphics[width=\linewidth]{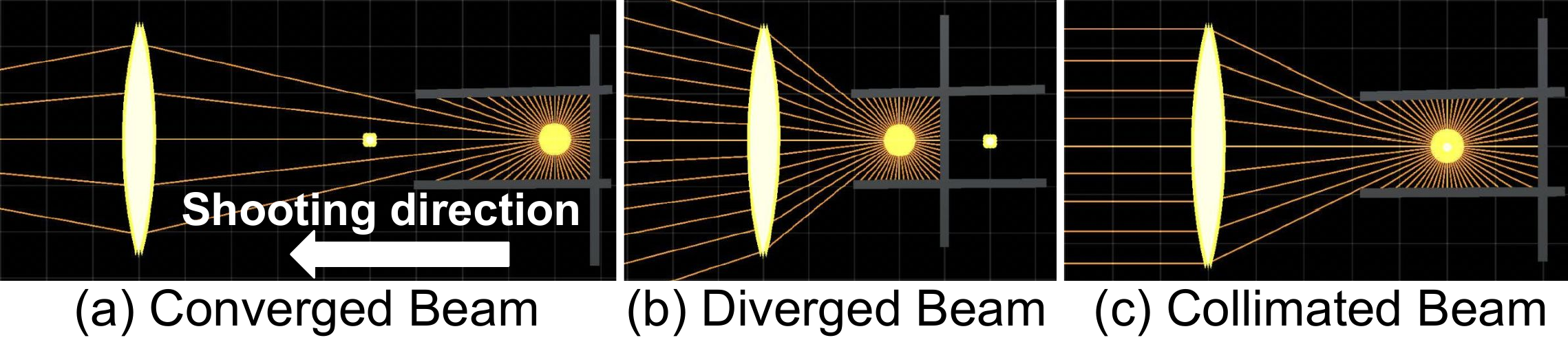}
\vspace{-0.1in}
\caption{
\newpart{
Illustration of the 3 types of laser beam shapes: converged, diverged, and collimated.
}
}
\label{fig:optics_exp}
\end{figure}

\newpart{
Furthermore, the improved optical design significantly affects the number and angle coverage of spoofable points as discussed in~\S\ref{sec:optical_setup}. Fig.~\ref{fig:optics_exp} illustrates the 3 types of laser beam shapes:  converged, diverged, and collimated. Among them, the collimated beam is the most ideal for LiDAR spoofing attacks because it can deliver the laser to the target with a minimum loss. To precisely calibrate the lens setup, we develop a device that can adjust the distance between the LD and the lens as designed. As shown in Fig.~\ref{fig:spoofer1}, the lens is connected to the frame with a hollow screw so that we can adjust it precisely. 
}

\nsection{Case Study Results on Specific LiDARs}
\label{appendix:specific-lidars}
\vspace{0.05in}

\nsubsection{The Use of Rare Laser Wavelength in OS1-32}

For OS1-32~\cite{OS1-32}, we find that we are not able to inject many points since we only have a 905 nm wavelength (SPL PL90\_3), which is also the setup for all prior works~\cite{shin2017illusion, cao2019adversarial, jiachen2020towards, hallyburton2022security, cao2023you}. %
The spoofing with 905 nm wavelength is not effective on OS1-32 that uses an 865 nm wavelength laser. 
We tried our best to purchase 865 nm laser diodes, but we find that high-power 865 nm laser diodes are actually not generally publicly available for personal use. We also tried low-power 860 nm laser diodes such as OPV380 and RLD85PZJ4, but they cannot work more than 4 W, which is too low to inject any points (our 905 nm diodes work at 90W to achieve successful injections).
From some perspective, this may imply that developing/using a LiDAR with a relatively rare wavelength may have a ``immunization'' effect against LiDAR spoofing attacks in practice since it may make it harder for attackers to acquire the corresponding attack laser diodes.
At this point, only a very small portion of LiDARs has this property; a recent survey~\cite{lidar_survey2021} reports that 905 nm wavelength laser is the most commonly used and 865 nm wavelength laser does not even appear in their graph due to its rarity.

Nevertheless, such a potential ``immunization'' effect would lose effectiveness if more LiDARs adopt 865 nm wavelength as this may boost the availability of 865 nm laser diodes on market. Also we find that not all LiDARs using a rare wavelength can directly have such an ``immunization'' effect. For example, although Realsense L515 is using 860 nm instead of 905 nm, we find that it is still directly attackable with the 905 nm attack laser since it does not have a band-pass filter to exclude light other than its wavelength.

\nsubsection{Relay Attack on Leddar Pixell}

\begin{figure}[h!]
\centering
\includegraphics[width=\linewidth]{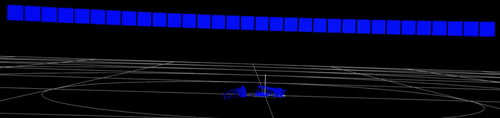}
\caption{Relay Attack on Leddar Pixell~\cite{pixell}. The attack moves the detected area to a much farther location. %
}
\label{fig:relay_attack}
\end{figure}

Flash LiDAR fires a wide diverging laser beam and needs to receive them at the same time. This mechanism is exploitable by the relay attack (\S\ref{sec:async_attack}), which is not effective on the scanning LiDARs because PD can only receive a very limited azimuthal range, and the spoofed points will always be located far from the spoofer as discussed in~\cite{shin2017illusion}. 
However, we find that the relay %
attack can be effective on the flash LiDAR as a removal attack. As its laser covers a wide range, the attacker can also disturb a wider range. If the attacker's laser is strong enough, they can remove the original points by moving them to farther points than the spoofer.
Fig.~\ref{fig:relay_attack} shows the results of the relay attack on Leddar Pixell~\cite{pixell}. The entire row of detection is moved to very far position.
Note that Leddar Pixell has a special design that consists of multiple rows of scanning. If it is a typical flash LiDAR, the relay attack can cause an effect on more rows. %

\begin{observation}{RQ2}
Compared to traditional scanning LiDARs, relay attack may be more effective on flash LiDARs as an object removal attack.
\end{observation}

\nsubsection{Zero-Distance Sensing of XT32}

As listed in Table~\ref{tbl:target_lidars}, XT32~\cite{XT32} is capable to measure the distance to LiDAR down to 0 m. The zero-distance sensing is not available in most LiDARs due to its technical challenges: detecting a very short time of laser flight is difficult to distinguish from noise due to unideal hardware calibration as discussed in~\cite{cao2023you}. However, it is technically realizable, which thus directly breaks the design assumption of the latest PRA attack~\cite{cao2023you} (i.e., needs enough MOT, \S\ref{sec:sync_removal_attack}).

\begin{observation}{RQ1}
LiDARs with zero-distance sensing do exist, which directly breaks the design assumption of the latest PRA attack~\cite{cao2023you}.
\end{observation}

\nsubsection{Extra Wide Vertical FOV of Helios 5515} \label{sec:case_helios}

\begin{figure}[t!]
    \begin{minipage}{.6\linewidth}
        \centering
        \includegraphics[width=\linewidth]{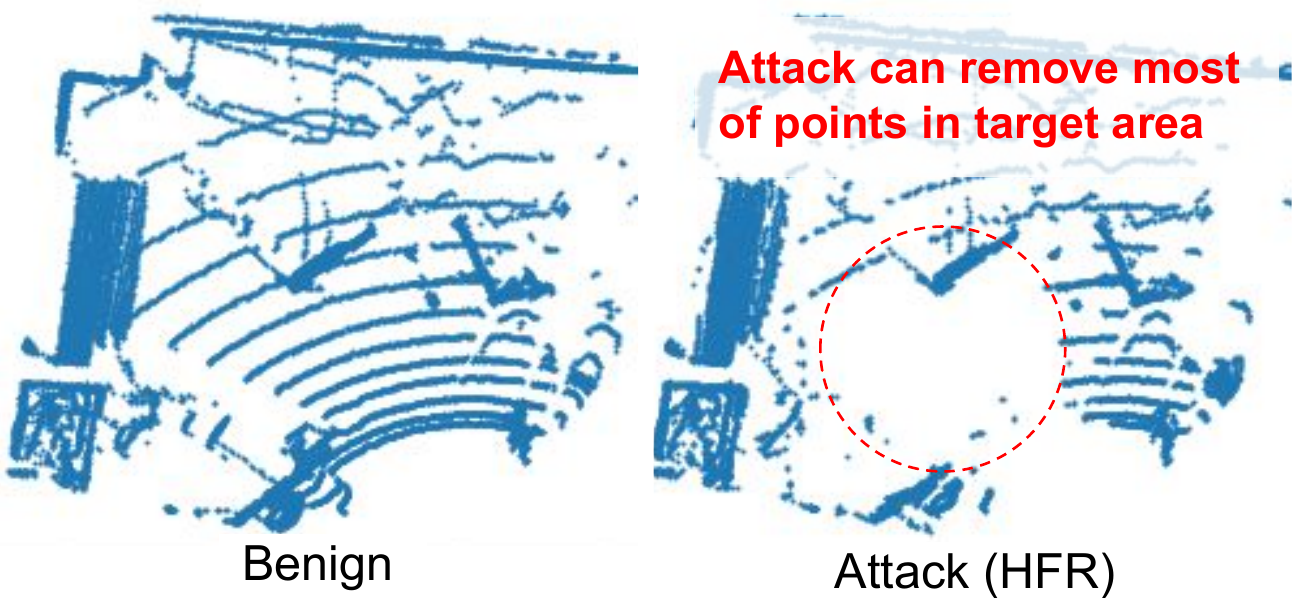}
        \caption{HFR attack effect on Helios~\cite{Helios}. Our spoofer cannot cover the entire FOV, but can still be effective in the center area.}
        \label{fig:helios_hfa}
   \end{minipage}\hspace{0.03in}
    \begin{minipage}{.37\linewidth}
        \centering
        \vspace{-0.1in}
        \includegraphics[width=\linewidth]{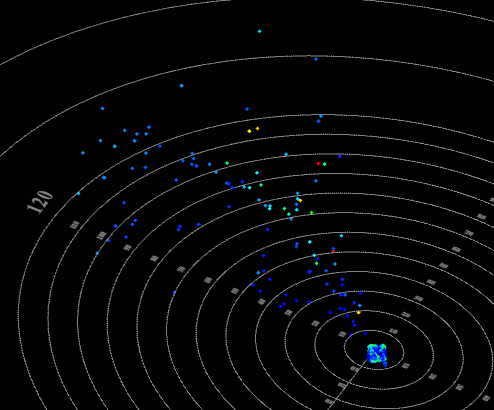}
        \caption{Spoofed points on XT32~\cite{XT32}.}
        \label{fig:xt32_100pt_injection}
    \end{minipage}
\end{figure}

As listed in Table~\ref{tbl:target_lidars}, Helios~\cite{Helios} has an extra wide vertical field-of-view (FOV), 70$^{\circ}$, which is far wider than the normal range of 30-40$^{\circ}$ (Table~\ref{tbl:target_lidars}).
It results in the lower attack success rate $\mathcal{R}=19.4\%$ even though the number of points is large, 3,203 points.
This is because our spoofer cannot cover all altitudes of Helios. However, it does not mean that Helios is more robust against attacks because the attacker does not need to attack the entire vertical FOV. Fig.~\ref{fig:helios_hfa} shows an example of the HFR attack effect on Helios. As shown, the HFR attack is successful in the middle of the vertical FOV, which can allow attackers to hide objects there. %

\nsubsection{Simultaneous Firing on OS1-32 and VLS-128}

\begin{figure}[h!]
\centering
\includegraphics[width=\linewidth]{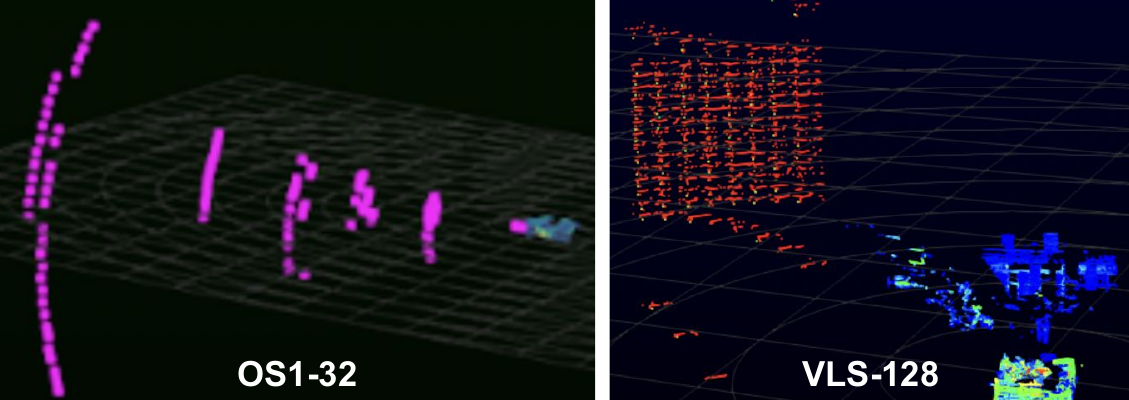}
\caption{Spoofing results for OS1-32 and VLS-128.
}
\label{fig:sim_firing}
\end{figure}

Fig.~\ref{fig:sim_firing} shows the spoofed points on OS1-32~\cite{OS1-32} and VLS-128~\cite{VLS128}.%
As shown, OS1-32 fires and scans 32 lasers vertically, and thus we can only move the depth of each vertical line with 32 points simultaneously.
VLS-128 shoots 8 lasers based on a predefined pattern, and thus we can only simultaneously change the depth of each group consisting of 8 points.
As we do not have the capability to selectively return a laser to each simultaneous laser, we are not able to achieve the CPI attack capability as discussed in~\S\ref{sec:sec_enchance_feats}.

\nsection{Taxonomy of 3D Object Detectors}
\label{appndix:obj_detector}

\textbf{Voxel-based Methods.}
Voxel-based method is a very early, but still dominant approach~\cite{lang2019pointpillars, zhou2018voxelnet, shi2020points, yan2018second}. To deal with the irregular structure of point clouds, this approach aggregates points into 3D voxels to make the CNN work effectively. 
PointPillars~\cite{lang2019pointpillars} is the most widely used in autonomous driving systems such as Baidu Apollo~\cite{apollo} and Autoware~\cite{autoware} because it can achieve higher throughput by constructing voxels as pillars perpendicular to the grounds although the z-axis resolution becomes coarse. In~\S\ref{sec:od_injection}, we find that the voxel-based method is slightly less robust than the other methods to the injection attacks as it cannot recognize the detailed geometry of points.
This design allows to achieve of higher throughput and is widely used in autonomous driving systems such as Baidu Apollo~\cite{apollo} and Autoware~\cite{autoware} since the z-axis of an object is typically unimportant in ground vehicles. Part-A$^2$~\cite{shi2020points} achieves a higher attack success rate than the above methods by leveraging multi-scale voxelization and anchor-based strategies.

\textbf{Point-based Methods.}
Point-based methods directly handle point cloud without voxelization. To efficiently handle point clouds, this approach utilizes permutation-invariant operators. PointRCNN~\cite{shi2019pointrcnn} is a two-stage method inspired by FastRCNN~\cite{girshick2015fast} in 2D object detection. PointRCNN generate 3D proposals and point features with PointNet backborns~\cite{qi2017pointnet++}.  3DSSD~\cite{yang20203dssd} is a single stage method inspired by SSD~\cite{liu2016ssd} in 2D object detection. By leaving only representative points in the downsampling, 3DSSD removes the feature propagation layer and achieves higher throughput than two-stage method.

\textbf{Point Voxel-based Methods.}
Point voxel-based method~\cite{shi2020pv, chen2019fast} is a hybrid approach of the voxel-based and point-based  methods. PV-RCNN~\cite{shi2020pv} is a two-stage method that uses the voxel-based method in the first stage (3D proposal generation) and the point-based method in the second stage (regional refinement). In~\S\ref{sec:od_injection}, PV-RCNN~\cite{shi2020pv} shows the highest robustness to the injection attacks. The second stage with PointNet backbone, which is not in 3DSSD, enables to obtain more detailed point geometry.

\begin{figure}[t!]
\centering
\includegraphics[width=0.8\linewidth]{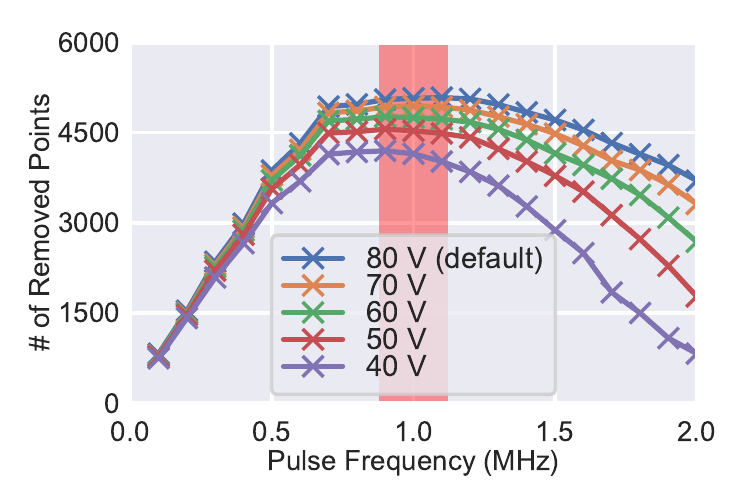}
\vspace{-0.1in}
\caption{The relationship between the attack pulse frequency and the removed points by HFR attack. 
}
\label{fig:relationship_freq_pts}
\end{figure}

\nsection{Impact of Pulse Frequency and Laser Drive Voltage on HFR Attack} \label{appndix:hfr_freq}
For HFR attack, the higher the attack pulse frequency is, the more effective the point removal attack capability should be since this can make it more likely for the attack pulse to (1) hit the laser receiving window on the victim LiDAR side to affect the legitimate point measurement; and (2) push the attack-induced random position measurements to be within MOT (and thus become undetectable).
However, highly-frequent laser firing makes the temperature of LD high and thus results in the degradation of laser intensity, which affects the spoofing capability.
We thus experimentally study this trade-off. As shown in Fig.~\ref{fig:relationship_freq_pts}, when we increase the frequency, the number of removed points peaks at $\sim$1 MHz and decreases after that. Thus, we use 1 MHz as the default attack laser frequency in our other experiments.

Meanwhile, the attack laser intensity at the firing time also practically affects the attack effectiveness, since a lower one is not able to ensure that the attack laser received at the victim is stronger than the legitimate one. To understand this, we also vary the attack laser drive voltage in our experiments. Since VLP-16 has $\sim$30V laser drive voltage, we vary the attack laser drive voltage from 40 to 80 V (80V is the maximum possible one in our setup). As shown in Fig.~\ref{fig:relationship_freq_pts}, the number of removed points monotonically decreases with the voltage value. Thus, we use 80V as the default voltage in our experiments.

\nsection{Criteria to Count the Number of Removed and Injected Points} \label{appndix:count_method}
To quantitatively evaluate the attack performance, we design a systematic method to count the number of removed and injected points by attacks. We conduct all indoor experiments 5 m away facing the wall of the room, i.e., the legitimate reflection should be from 5 m away.
For point injection attacks, we judge whether a point is spoofed or not based on its intensity. When the intensity scale is 0 to 255, The intensity of reflections from the walls of the room is typically below 70. On the other hand, the attack laser has more than 200 as we directly shoot lasers at the LiDAR without any reflections.
We thus used 80 as the threshold for indoor experiments to differentiate whether legitimate or injected points. For the outdoor experiments, we selected an adequate threshold from 80 to 150 based on the maximum intensity in a benign point cloud.
For point removal attacks, we also utilize the intensity of the point. We first identify the attack-induced spoofed points with the same threshold of intensity. Next, we subtract the remaining points in the attacked point cloud from the benign point cloud; now the remaining points in the benign point cloud are thus the removed benign points by the attack. To calculate the point removal percentage of each azimuth, we split the area into pies corresponding to each azimuth and apply the same calculation.
Based on our measurement, our method can capture 96\% of injected or removed points, i.e., 4\% of injected or removed points could be missed.

\end{document}